\PassOptionsToPackage{dvipsnames}{xcolor}
\documentclass[11pt,letterpaper]{article}
\usepackage[margin=1in]{geometry}
\usepackage{setspace}

\usepackage{tikz}
\usetikzlibrary{arrows}
\usepackage{filecontents}

\usepackage{mathtools}

\usepackage{subfigure}
\usepackage{float}
\usepackage{caption}
\usepackage{graphicx}
\usepackage{amsmath,amsthm,amssymb,amsfonts}
\usepackage{algorithm}
\usepackage{listings}
\usepackage[colorlinks,linkcolor=blue]{hyperref}
\usepackage{url}

\usepackage{colortbl}

\usepackage{pdfpages}

\usepackage{authblk}

\newtheorem{theorem}{Theorem}
\newtheorem{lemma}{Lemma}
\newtheorem{definition}{Definition}
\newtheorem{remark}{Remark}

\newcommand{\dx}{\textrm{\,d}x}
\newcommand{\dt}{\textrm{\,d}t}

\newcommand{\mynote}[3]{
    \fbox{\bfseries\sffamily\scriptsize#1}
{\small$\blacktriangleright$\textsf{\emph{\color{#3}{#2}}}$\blacktriangleleft$}}
\newcommand{\va}[1]{\mynote{Vitaly}{#1}{magenta}}

\renewcommand{\le}{\leqslant}  
\renewcommand{\leq}{\leqslant}  
\renewcommand{\geq}{\geqslant}  

\DeclarePairedDelimiter{\ceil}{\lceil}{\rceil}
\DeclarePairedDelimiter{\floor}{\lfloor}{\rfloor}

\author{
Alexander Slastin, ITMO University \and
Dan Alistarh, IST Austria \and
Vitaly Aksenov, City, University of London
}

\title{Efficient Self-Adjusting Search Trees via Lazy Updates}
\date{}

\begin{document}

\singlespacing

\maketitle

\begin{abstract}
    Self-adjusting data structures are a classic approach to adapting the complexity of operations to the data access distribution. 
    While several self-adjusting variants are known for both binary search trees and B-Trees, existing constructions come with limitations. 
    For instance, existing works on self-adjusting B-Trees do not provide static-optimality and tend to be complex and inefficient to implement in practice. 
    In this paper, we provide a new approach to build efficient self-adjusting search trees based on state-of-the-art non-adaptive structures. 
    We illustrate our approach to obtain a new efficient self-adjusting Interpolation Search Tree (IST) and B-Tree, as well as a new self-adjusting tree called the Log Tree. 
    Of note, our self-adjusting IST has expected complexity in $O(\log \frac{\log m}{\log ac(x)})$, where $m$ is the total number of requests and $ac(x)$ is the number of requests to key $x$. 
    Our technique leads to simple constructions with a reduced number of pointer manipulations: this improves cache efficiency and even allows an efficient concurrent implementation. 




\end{abstract}

\newpage

\section{Introduction}
\label{sec:introduction}

Whenever it is necessary to use efficient key-value sequential data structure with \texttt{get}, \texttt{insert}, and \texttt{delete} operations, the choice usually falls on hash tables or balanced search trees~\cite{knuth1997art} for their optimal amortized or worst-case asymptotic guarantees. These data structures assume that every element has the same probability to be accessed, in other words, the data access distribution is uniform. However, in many real workloads, the frequency of accesses to different elements is not uniform. This fact is well-known and is modeled in several industrial benchmarks, such as TPC-C~\cite{TPCC}, or YCSB~\cite{cooper2010benchmarking}, where the  access distributions are heavy-tailed, e.g., following a Zipf distribution~\cite{Zipf}. 
In that case, one would desire to use a sequential data structure that adjusts to the workload, i.e., more frequently accessed keys are accessed faster. To do this, the self-adjusting data structures were designed. The most well-known of them are Splay Trees~\cite{splay-tree} and Tango Trees~\cite{demaine2007dynamic}.
Indeed, these data structures provide strong performance on skewed Zipfian workloads in comparison to the standard binary search trees, such as Treaps, Red-black trees, or AVL trees.
Moreover, Splay Trees also provide strong theoretical guarantees, in the form of \emph{static-optimality}: its efficiency is similar to the efficiency of the best offline search tree for the given access sequence. Moreover, Tango Trees provide an even stronger guarantee, being $O(\log \log n)$ dynamic-optimal: its efficiency is $O(\log \log n)$ away from the efficiency of the best \emph{online}  binary search tree for the given access sequence.

However, in practice, e.g., for a database index, the engineers usually employ non-self-adjusting multiway trees such as B-Trees~\cite{b-tree} or Interpolation Search Trees (IST)~\cite{DynamicIST}. These approaches currently lack self-adjusting properties, such as static-optimality. 
Specifically, existing self-adjusting data structures such as the CBTree~\cite{CBTree}, the Splay Tree~\cite{splay-tree}, and multiway variations such as B-Tree variants~\cite{bose2008dynamic, demaine2021belga, k-splay-tree}, are inferior in efficiency to the mentioned above non-self-adjusting multiway trees even on moderately-skewed workloads. 

In this paper, we address this limitation by introducing a general approach to build efficient self-adjusting data structures from state-of-the-art non-self-adjusting structures. At a high level, our approach works by combining lazy rebuilding and access counters, constructing an  ideal static data structure from the list of (key, number of key accesses) pairs. 
Using this approach, we create \emph{self-adjusting} variants of well-known multiway data structures: Interpolation Search Tree (IST), B-Tree, as well as a new data structure called the Self-Adjusting Log Tree.
We show that the resulting data structures provide significant performance benefits over original, non-self-adjusting, tree designs, and come with theoretical guarantees such as static-optimality, for a large class of access distributions.
More precisely, our self-adjusting IST has expected complexity of $O(\log \frac{\log m}{\log ac(x)})$, where $m$ is the total number of requests and $ac(x)$ is the number of requests to key $x$.
As a nice additional result, our self-adjusting data structures support range queries with proven bounds, allowing us to get all keys from a required segment, and to efficiently calculate some functions on a range of keys.



\noindent\textbf{Related work.}
There exists a line of work on self-adjusting data structures based on binary search trees~\cite{splay-tree,buadoiu2004simplified,wang2006log} and on B-Trees~\cite{bose2008dynamic, demaine2021belga, k-splay-tree}.
Our work differs from them in the following three main properties.

At first, most of the papers about B-Trees discuss dynamic-optimality and they do not care about the static-optimality, thus, they do not provide the corresponding bounds. In this work, we present a self-adjusting B-Tree with the static-optimality property. Then, we go further and consider Interpolation Search Tree~\cite{DynamicIST} which has better  expected complexity $O(\log \log n)$ on search requests than other known tree data structures: we make its self-adjusting version that serves a request in $O(\log{\frac{\log{m}}{\log{ac(x)}}})$ where $m$ is the total number of requests and $ac(x)$ is the number of requests to key $x$. This complexity is an improvement to the static-optimality.

Second, our data structures are based on lazy rebuilding paradigm, i.e., we rebuild the whole subtree if the number of requests exceeds some bound. This leads to the following properties: 1)~the algorithms become much simpler to understand, 2)~it reduces the number of manipulations with pointers and, thus, improves cache-efficiency, and 3)~we can make a lock-free efficient concurrent version of our data structures using the approach in~\cite{CIST} (see Section~\ref{sec:concurrent}).

Finally, we improve the range queries on top of the self-adjusting data structures. In the previous works, range queries do not affect the accesses for all the keys in the range, while in our implementations we can increase the number of requests for each element so that all elements from that range will be accessed faster.

\noindent\textbf{Roadmap.}
In Section~\ref{sec:model}, we give the most important notions that we are going to use throughout the paper.
In Section~\ref{sec:gsat-approach}, we describe our lazy-rebuilding approach to create self-adjusting data structures and prove bounds for the operations on them.
In Section~\ref{sec:sait}, we show how to design self-adjusting IST and prove the most interesting expected complexity bound.
In Section~\ref{sec:sabt}, we apply our approach to B-Tree.
In Section~\ref{sec:salt}, we present a new self-adjusting Log Tree.
In Section~\ref{sec:range-queries}, we explain how to get range queries with nice bounds on our versions of data structures.
In Section~\ref{sec:concurrent}, we briefly show how to make lock-free concurrent versions of our data structures.
In Section~\ref{sec:experiments}, we present that our implementations, typically, work better on read-only and lazy-delete workloads than their standard counterparts.
Finally, we conclude with Section~\ref{sec:conclusion}.

\section{Overview of Self-Adjusting Maps}
\label{sec:model}

In this section, we present the general properties of our target data structures.

\subsection{The Map Definition. }
\label{subsec:map-def}

Our data structures provide the following operations, i.e., \texttt{Map} interface:
\begin{enumerate}
    \item \texttt{get(key)}~--- returns a value corresponding to the \texttt{key}, if the \texttt{key} exists in the data structure and non-deleted, or \texttt{null}, otherwise.
    \item \texttt{insert(key, value)}~--- adds the pair of a \texttt{key} and a \texttt{value} to the data structure if the \texttt{key} is not already present.
    \item \texttt{delete(key)}~--- removes the \texttt{key} from the data structure if it exists.
\end{enumerate}


As suggested, in our self-adjusting data structures, the \texttt{delete} operation does not physically delete the \texttt{(key, value)} pair~--- instead, it just marks it as deleted. 


\subsection{Static-Optimality}
\label{subsec:static-optimality}

One of the important properties required from self-adjusting data structures is the static-optimality.

\begin{theorem}
\label{th:static-optimality}
If each element is accessed at least once and the requests are known in advance, then the total access time in the static optimal search tree is $\Theta\Big(m + \sum\limits_{i=1}^n ac(i) \times \log{\Big(\frac{m}{ac(i)}\Big)}\Big)$, where  $n$ is the number of pairs in the data structure, $ac(i)$ is the number of accesses made to the $i$-th element and $m$ is the total number of accesses to all elements $(m = \sum\limits_{m=1}^n ac(i))$.

If a dynamic tree data structure serves online requests with this complexity we say it has the \emph{static-optimality} property.
\end{theorem}

This theorem holds due to the information theory results~\cite{splay-tree} since search trees use only comparisons.
We want our data structures to hold this static-optimality property or provide even better complexity.

The most renowned tree with this property is Splay Tree~\cite{splay-tree}. However, there is a line of work presenting data structures based on binary search trees~\cite{demaine2007dynamic,buadoiu2004simplified,wang2006log} and on B-Trees~\cite{bose2008dynamic, demaine2021belga, k-splay-tree}.

\subsection{Target data structures}
\label{subsec:target-ds}

In Section~\ref{sec:gsat-approach}, we present an approach that can make a self-adjusting tree from a version with the lazy rebuild, i.e., when a subtree with a large number of requests is rebuilt into its ideal version, with the required depth guarantees. We applied this approach to two well-known data structures: Interpolation Search Tree (Section~\ref{sec:sait}) and B-Tree (Section~\ref{sec:sabt}).

\noindent\textbf{The Interpolation Search Tree (IST). }
The interpolation search tree was proposed in~\cite{DynamicIST}. Its worst-case amortized bounds for all operations are $O(\log^2{n})$, but for a wide class of distributions called \texttt{smooth}~(Definition~\ref{def:smooth}), the expected time of all operations is $O(\log{\log{n}})$. This is achieved by using $\sqrt{n}$ children in the root instead of the fixed number and array $ID[1 \ldots \ceil{n^{\alpha}}]$ with $\alpha \in [\frac{1}{2}, 1)$ which interpolates the requested key and finds a proper subtree in $O(1)$ time.

\noindent\textbf{B-Tree. }
The B-Tree is a balanced search tree whose nodes have the number of keys from $B$ to $2 \cdot B$ and the number of children from $B + 1$ to $2 \cdot B + 1$, respectively, where $B$ is a preselected constant. 

All operations in the B-Tree search for a node with the specified key. To do that, they traverse down the tree, starting from the root node, and use the binary search in each node: 1)~to stop traversal, if the requested key was found, or 2)~to find the proper child node to look at.

This data structure is usually used as a database index due to good work with memory and $O(\log{n})$ time complexity.

In this work, as a basis of our self-adjusting version, we consider a lazy B-Tree that is balanced by rebuilding subtrees from scratch.

\subsection{Range queries}
\label{subsec:rq-def}

The existing self-adjusting data structures are usually not supplemented with range queries. 
In this work, we require our data structures to support the following range queries, denoting the key as $x_i$ and the corresponding value as $y_i$:
\begin{enumerate}
    \item \texttt{get(a, b)}~--- returns an array of keys that belong to the range $[a, b]$ and currently present in the tree: $[ x_i < x_{i + 1} < \ldots < x_{i + l} ] \subseteq [a, b]$;
    \item \texttt{calculate(a, b)}~--- returns the result of applying function $\odot$ to all values in the order of keys, which are currently present in the tree: $y_i \odot y_{i + 1} \odot \ldots \odot y_{i + l} $ for which $[ x_i < x_{i + 1} < \ldots < x_{i + l} ] \subseteq [a, b]$;
    \item \texttt{update(a, b, c)}~--- applies to all values whose keys are present and belong to the range $[a, b]$ the following action: $y_i' = y_i \star c$.
\end{enumerate}

In Section~\ref{sec:range-queries}, we explain how to change the newly designed self-adjusting data structures, so that they support range queries.


\section{Generic Self-Adjusting Search Trees}
\label{sec:gsat-approach}

In this section, we prove the common bounds on time and memory complexity for our self-adjusting IST and B-Tree. Thus, we refer now to their generic representation: a generic self-adjusting tree.

\subsection{Tree Structure}
\label{subsec:tree-structure}

Generic Self-Adjusting Tree (GSAT) is a multiway tree, for which the degree of a node depends on the number of requests to the underlying set. More precisely, the degree of the root is $O(\operatorname{D}(m))$, where $\operatorname{D}$ is the \emph{degree} function, i.e., the maximum number of internal keys inside the node, and $m$ is the total number of requests to all the elements in the tree. This is the major difference between our data structures and the standard ones where the degree function depends on the set size. Additionally, we have a function $\operatorname{S}(T, key)$ which is used to search for the proper children subtree of the node (tree) $T$ that should store $key$. In \emph{ideal} GSAT, subtrees on the first level of the tree have an almost equal total number of requests to all their elements, i.e., not exceeding $\frac{m}{\operatorname{D}(m) + 1}$, and hence a child of the root of an ideal GSAT has a degree $O(\operatorname{D}(\frac{m}{\operatorname{D}(m) + 1}))$.

\subsection{Formal definitions}
\label{subsec:formal-defs}

For simplicity, we suppose that all keys in this work are integers.

Let $a$ and $b$ be some integers with $a < b$. A Generic Self-Adjusting Tree (GSAT) with boundaries $a$ and $b$ for a set $X = \{ x_1 < x_2 < \ldots < x_n \} \subseteq [a, b]$ of $n$ elements with $ac_1, ac_2, \ldots, ac_n$ accesses made to respective elements, consists of:

\begin{enumerate}
\itemsep0em 
    \item An integer $m = \sum_{i=1}^n ac_i$~--- the total number of accesses to the elements of the set $X$.
    \item An array $REP[1 \ldots k]$ of representatives $x_{i_1},x_{i_2}, \ldots ,x_{i_k}$, where \\ $i_1 < i_2 < \ldots < i_k$ and $REP[j] = x_{i_j}$. Furthermore, $k$ does not exceed $\ceil{\operatorname{D}(m)}$.
    \item An array $AC[1 \ldots k]$ of the number of accesses made to $x_{i_1},x_{i_2}, \ldots ,x_{i_k}$, respectively.
    \item The subtrees of the root are GSATs $T_1, T_2, \ldots ,T_{k + 1}$ for the subsets $X_1, X_2, \ldots , X_{k + 1}$, where \\ $X_1 = \{ x_1, \ldots , x_{i_1 - 1} \}$, $X_j = \{ x_{i_{j - 1} + 1}, \ldots , x_{i_j - 1} \}$ for $2 \le j \le k$,  $X_{k + 1} = \{ x_{i_k + 1}, \ldots , x_n \}$. Furthermore, $T_1$ has boundaries $a$ and $x_{i_1}$, $T_j$ has boundaries $x_{i_j - 1}$ and $x_{i_j}$ for $2 \le j \le k$, $T_{k + 1}$ has boundaries $x_k$ and $b$.
    \item A function $\operatorname{S}(T, key)$ that returns the index $i$ such that either $REP[i]$ is equal to $key$ or $T_i$ should contain $key$, or \texttt{null} if there is no such $T_i$.
\end{enumerate}

The array $REP$ contains several keys from the set $X$. In ideal GSATs we require these keys to be equally spaced in accordance with the number of accesses to the elements between them. Let us denote $m(T_i)$ for the number of accesses made to all elements of a subtree $T_i$.

\begin{definition}
\label{def:ideal-gsat}
A GSAT is \emph{ideal} if $m(T_j) \leq \frac{m}{\operatorname{D}(m) + 1}$ for all $j$ and the GSATs $T_1, T_2, \ldots, T_{k + 1}$ are also \emph{ideal}.
\end{definition}

\begin{figure}[H]
  \centering
  \caption{General GSAT structure. The upper part of the red square corresponds to the segment on which the GSAT was built, and its lower part is equal to the total number of accesses to the elements of the tree.
  }
  \label{pic:gsat}
  \includegraphics[width=\linewidth]{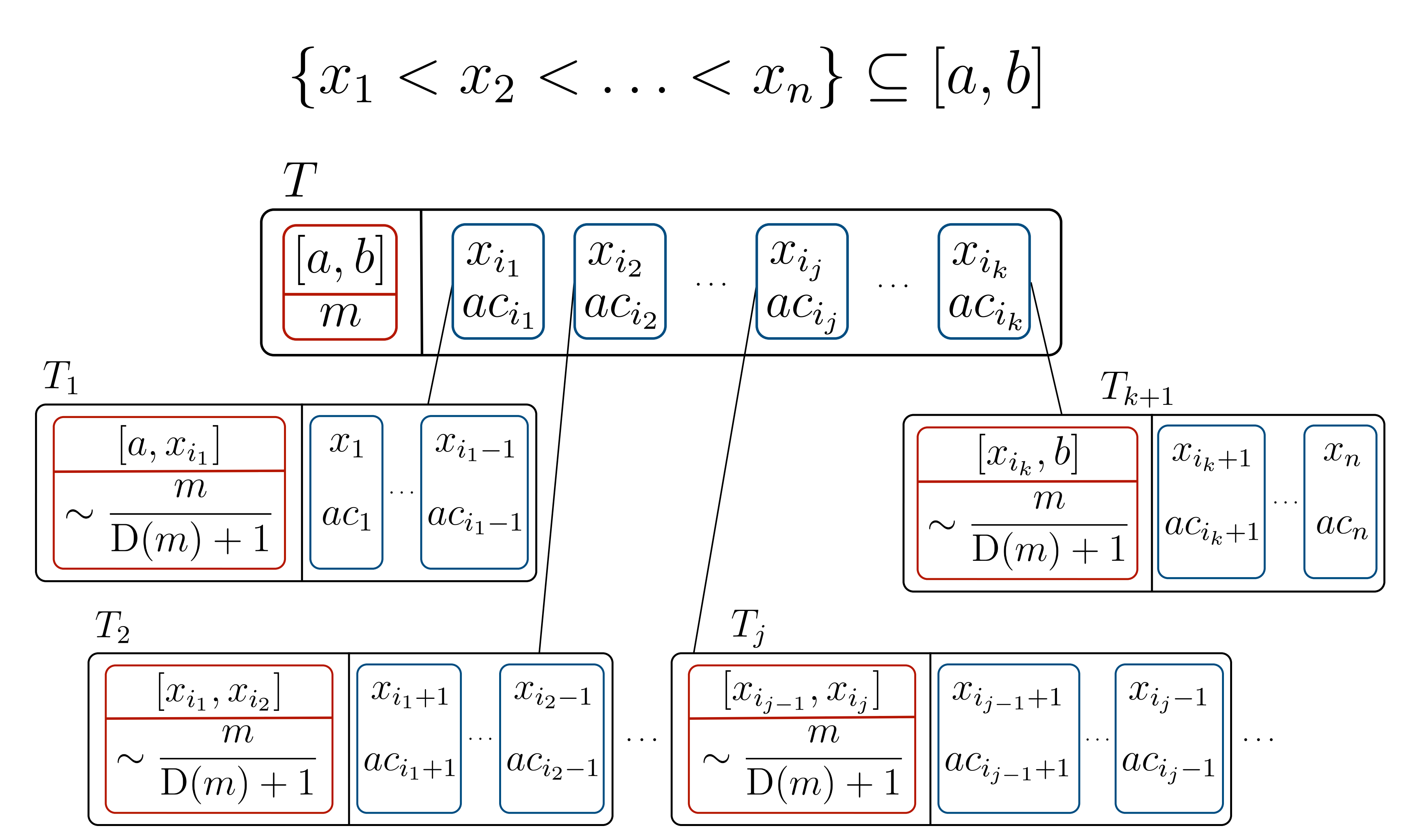}
\end{figure}

The general structure of the GSAT is shown on Figure~\ref{pic:gsat}.

In our work, we require the degree function $\operatorname{D}(m)$ to be \emph{sqrt-bounded}:
\begin{definition}
\label{def:sqrt-bounded}
The function $f(m)$ is \emph{sqrt-bounded} if:
$1 \leq f(m) \leq \sqrt{m}$ except for the set of a finite measure on which $f(m)$ is bounded.
\end{definition}

Depending on ${\operatorname{D}(m)}$ and $\operatorname{S}(T, key)$, the time complexities of operations as well as memory consumption may vary. For further analysis, we consider two variations of these parameters for our trees:
\begin{enumerate}
    \item ${\operatorname{D}(m)} = \sqrt{m}$. $\operatorname{S}(T, key)$ uses the interpolation search, based on the special array $ID[1 \ldots m^{\alpha}]$ with $\alpha \in [\frac{1}{2}, 1)$, and, then, the exponential search, together with binary search. We call this parameterization a Self-Adjusting Interpolation Search Tree (SAIT). SAIT in our scheme is similar to the standard Interpolation Search Tree~\cite{DynamicIST}. However, it is based on the number of accesses rather than on the number of elements.
    \item ${\operatorname{D}(m)} = B$, where $B$ is a preselected constant, and $\operatorname{S}(node, key)$ uses binary search. We call this parameterization a Self-Adjusting B-Tree (SABT).
\end{enumerate}

\subsection{Construction of an Ideal Tree}
\label{subsec:construction}

We start with an efficient algorithm to build the ideal GSAT given an ordered array of elements and their number of accesses. To start with, we count prefix sums of all accesses made to an ordered array. Then, we generate representatives in the root node one after another using the binary search on the collected prefix sums, splitting the number of accesses almost evenly between the subtrees. The overall pseudocode is present in Appendix~\ref{app:pseudocode-ideal}. 

Each representative is found using the binary search and the subtrees are built recursively. For $j = 1 \ldots |REP|$, $m(T_j) < \ceil[\Big]{\frac{m}{\ceil{\operatorname{D}(m)} + 1}}$ and $m(T_{|REP| + 1}) \leq m - \ceil{\operatorname{D}(m)} \times \ceil[\Big]{\frac{m}{\ceil{\operatorname{D}(m)} + 1}} \leq m - \frac{\ceil{\operatorname{D}(m)} \times m}{\ceil{\operatorname{D}(m)} + 1} = \frac{m}{\ceil{\operatorname{D}(m)} + 1}$, so, each subtree has no more than $\frac{m}{\operatorname{D}(m) + 1}$ accesses in total. Also, array $REP$ can have less than $\operatorname{D}(m)$ representatives as shown on Figure~\ref{pic:non-full-gsat}:

\begin{figure}[H]
  \centering
  \caption{Non-full GSAT. $\operatorname{D}(m) = \sqrt{m}$, $x = [1, 2, 3, 4] \subseteq [1, 5]$, $ac = [1, 18, 2, 3]$.}
  \label{pic:non-full-gsat}
  \includegraphics[width=\linewidth]{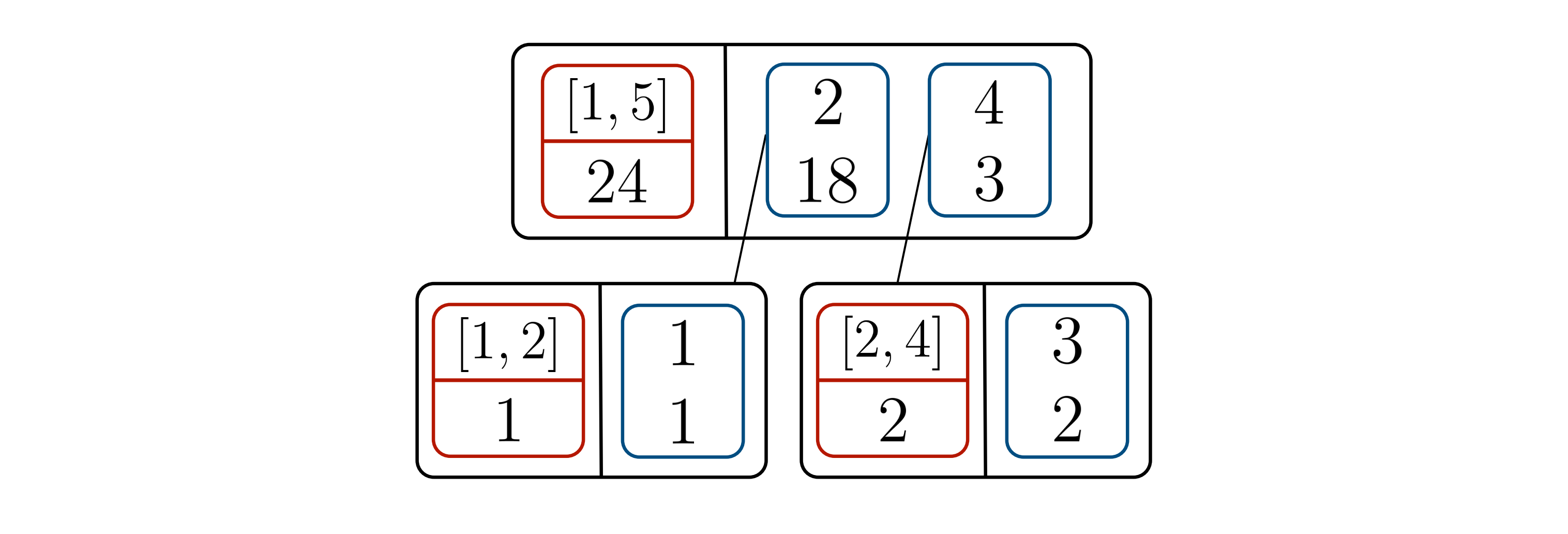}
\end{figure}

Please note that for SAIT we have to build an array $ID$ of size $m^\alpha$, $\alpha \in [\frac{1}{2}, 1)$. It can be built after all the representatives and subtrees are chosen, so, the analysis for building this array can be done separately from the analysis for building the representatives and subtrees.

Let us prove the several properties of that algorithm and the ideal GSAT.

\begin{theorem}
\label{th:gsat-construction}
An ideal GSAT for a set of size $n$ and the total number of requests $m$ can be built in $O(m)$ time and requires $O(m)$ memory. Also, GSAT has depth $O(\log{m})$.
\end{theorem}
\begin{proof}
Let $Time(m)$ be the time required to build a GSAT, then $$Time(m) \le c_1 \times \operatorname{D}(m) + \log_2(n) \times \operatorname{D}(m) + (\operatorname{D}(m)+ 1) \times Time\Big(\frac{m}{\operatorname{D}(m) + 1}\Big):$$
1)~$c_1 \times \operatorname{D}(m)$ is spent to build the node and additional data structures, e.g., a segment tree for range queries (however, $ID$ array is not considered here, and its complexity is calculated separately in Theorem~\ref{th:sait-construction});
2)~$\log_2(n) \times \operatorname{D}(m)$ is spent building representatives array in the root via binary search since the total number of keys is $n$;
3)~there are no more than $(\operatorname{D}(m)+ 1)$ subtrees $T_i$ for which $m(T_i) \leq \frac{m}{\operatorname{D}(m) + 1}$. Finally, $n \leq m$, so $Time(m) \leq (c_1 + \log_2(m)) \times \operatorname{D}(m) + (\operatorname{D}(m)+ 1) \times Time\Big(\frac{m}{\operatorname{D}(m) + 1}\Big)$. We prove that this function is $O(m)$ in Lemma~\ref{lem:gsat-reccurence}.

In the same manner, we can bound the required memory $Mem(m)$, $$Mem(m) = c_2 \times \operatorname{D}(m) + (\operatorname{D}(m)+ 1) \times Mem\Big(\frac{m}{\operatorname{D}(m) + 1}\Big) = O(Time(m)) = O(m)$$

Finally, let $d(m)$ be the depth of the ideal GSAT. Then, $d(m) = 1 + d\Big(\frac{m}{\operatorname{D}(m) + 1}\Big)$. $\operatorname{D}(m)$ is sqrt-bounded function, so $\operatorname{D}(m)$ is at least one and $\frac{m}{\operatorname{D}(m) + 1} \leq \frac{m}{2}$, thus, GSAT has depth $O(\log m)$. 
\end{proof}

\subsection{Worst-case guarantees}

\label{subsubsec:map-ops}

\texttt{insert}, \texttt{delete}, and \texttt{get} operations for our trees are implemented in the way of~\cite{DynamicIST}, using an approach based on the counters to amortize the tree rebuilding cost. 
We associate with each node $v$ a counter $C(v)$. It counts the number of operations done in subtree $T_v$ with the root $v$. Insertion creates a new node for the requested key and links it to the parent node if the tree does not already contain the key. Deletion is performed by \emph{marking} the requested key as deleted.

During any operation for some key, we increment counters of each node on the traversed path, until the key is found or there is no such key. Then, for a node $u$ among the traversed ones, which has the \emph{minimum} depth and whose counter overflows, we rebuild the whole subtree $T_u$ into \emph{ideal} GSAT for the set of \emph{unmarked} keys stored in $T_u$ and initialize the counters of all nodes of the new subtree $T_u$ to zero. Counter $C(v)$ \emph{overflows}, whenever its value exceeds $\frac{im(T_v)}{4}$, where $im(T_v)$ is $m(T_v)$ when $T_v$ was rebuilt for the last time.
Their pseudocode can be found in Appendix~\ref{app:pseudocode-operations}.

Now, we prove the worst-case running time of these operations.

\begin{lemma}
\label{lem:gsat-total-depth}
The worst-case depth of GSAT with $m$ total number of accesses, after executing \texttt{insert}, \texttt{delete}, \texttt{get} operations, is $O(\log m)$.
\end{lemma}
\begin{proof}
Consider a node $v$ and its parent $w$. Then $m(T_v) \leq \frac{im(T_w)}{4} + \frac{im(T_w)}{\operatorname{D}(m) + 1}$, since no more than $\frac{im(T_w)}{4}$ accesses could occurred in $T_v$, otherwise, $T_w$ would have been rebuilt. Also $\frac{im(T_w)}{\operatorname{D}(m) + 1} \leq \frac{im(T_w)}{2}$, so $m(T_v) \leq \frac{im(T_w)}{4} + \frac{im(T_w)}{2} = \frac{3}{4} \times im(T_w)$, thus, the depth of GSAT is $O(\log im(T_{root}))$. Since $m(T_{root}) = m(T) = m  \leq im(T) + \frac{im(T)}{4} = \frac{5}{4} \times im(T)$, we conclude that the depth is $O(\log m)$.
\end{proof}

\begin{lemma}
\label{lem:gsat-rough-depth}
The amortized cost of \texttt{insert}, \texttt{delete}, or \texttt{get} operations in GSAT (not counting the time for the search in the node) is $O(\log m)$, that is, the total cost of the first $m$ operations is $O(m \times \log{m})$.
\end{lemma}
\begin{proof}
We use the following accounting scheme to amortize the rebuild over all operations: each operation puts one token on each node on the path to the target element. A token represents the ability to pay for $O(1)$ computation time. By Lemma~\ref{lem:gsat-total-depth} each operation adds no more than $O(\log m)$ tokens. Rebuilding $T_v$ costs $O(m(T_v))$. From properties of the counter we get $m(T_v) \leq \frac{5}{4} \times im(T_v)$ and $C(v) \geq \frac{im(T_v)}{4}$, thus, we have enough tokens in the subtree to pay for its rebuild.
\end{proof}

When rebuilding a GSAT, the number of requests to an element affects its depth in the new tree, that is, the greater the number of requests, the smaller its depth will be, and the main drawback of the previous Lemma~\ref{lem:gsat-rough-depth} is that we do not take this relationship into account in any way. The following lemma allows us to correct this problem and show the static-optimality property of our trees.

\begin{lemma}
\label{lem:gsat-nice-depth}
The amortized cost of an \texttt{insert}, \texttt{delete}, or \texttt{get} operation in GSAT (not counting the time of the search in the node) for an element $x$ is $O(\log_2 \frac{m}{ac(x)} )$, where $ac(x)$ is the number of accesses to $x$.
\end{lemma}
\begin{proof}
Let $x$ be the element with $ac(x)$ accesses which has depth $d$ after the last rebuild of the subtree, containing key $x$. Let $T^{(k)}$ denote the subtree whose root has depth $k$ (root has depth $0$).

\va{update a little}
At any moment the following inequalities hold by the counter properties: $$m(T^{(1)}) \leq im(T^{(1)}) + \frac{m(T^{(0)})}{4} \leq \frac{m}{\operatorname{D}(m) + 1} + \frac{m}{4} \leq \frac{3}{4} \times m = \frac{m}{A} \text{ with } A = \frac{4}{3},$$ $$ m(T^{(i)}) \leq im(T^{(i)}) + \frac{m(T^{(i - 1)})}{4} \leq \frac{3}{4} \times m(T^{(i - 1)}) \leq \frac{m}{A^i}.$$ So for $T^{(d)}$, which contains the key $x$ inside its root node, we have: $\frac{m}{A^d} \geq m(T^{(d)}) \geq ac(x)$ that implies $d \leq \log_A\Big(\frac{m}{ac(x)}\Big)$, so, key $x$ has depth $O(\log_2(\frac{m}{ac(x)}))$.

The rebuilding is again simply amortized over the number of updated counters as in Lemma~\ref{lem:gsat-rough-depth}.
\end{proof}

Now, we need to make an important remark.
\begin{remark}
In this paper, we consider $ac(x)$ to be the number of accesses to the key $x$ in the version of the tree at the moment of an operation. Note, that logically deleted keys are removed after the rebuilding, thus, losing all the accesses. So, we get the static-optimality only for workloads without deletions.
\end{remark}

\subsection{Expected time analysis}

We calculate the expected complexities of operations to improve the worst-case bounds, if the keys are requested in accordance with some discrete distribution.
For example, we can improve the bound in Lemma~\ref{lem:gsat-nice-depth} to have $\log_B$ instead of $\log_2$ for B-Tree.

Let $p$ be the probability mass function on integers lying $[a, b]$. Let $F_{n}$ be a random sequence of size $n$ which is generated by drawing independently $n$ keys according to $p$. 

\begin{definition}
\label{def:random-gsat}
A $p[a, b]$-random GSAT is a GSAT with boundaries $[a, b]$ with $n$ elements and $m$ total number of accesses to all its elements, generated by the following actions, provided that $p$ is a probability mass function for which $\Big(\sum\limits_{x=a}^{b - 1} p(x) = 1\Big)$:
\begin{enumerate}
\item Take a random sequence $F_{n'}$ and build an ideal GSAT from its content, treating the number of occurrences of the element in the file as the number of accesses to it.
\item Perform a sequence of $p$-random operations: $i$ $p$-random \texttt{insert}, $d$ $p$-random \texttt{delete} and $g$ $p$-random \texttt{get} $Op_1, \ldots, Op_{i + d + g}$ so $i + d + g = m$ and $i - d = n$. An operation is $p$-random if it operates with a random key drawn according to $p$.
\end{enumerate}
\end{definition}

An ideal GSAT has already the best bounds for its depth, i.e., $O(\log \log m)$ for SAIT. Now, we want to prove that with high probability a random GSAT has the same depth bounds until the next rebuild of the entire tree. The following lemmas in this sub-section together show this fact.

\begin{lemma} 
\label{lem:subtree-prob}
Let $T$ be a $p[a, b]$-random GSAT and let $T'$ be a subtree of $T$. Then there are integers $c$, $d$ such that $T'$ is a $p'[c, d]$-random GSAT such that $a \leq c < d \leq b$ and for $x = c, c + 1, \ldots, d - 1$: $$p'(x) = \frac{p(x)}{\sum\limits_{y=c}^{d - 1} p(y)}.$$
\end{lemma}
This lemma is proven in the same way as in~\cite[Lemma 4]{DynamicIST}.

\begin{lemma}
\label{lem:prob-size}
Let $T$ be a $p[a, b]$-random GSAT with the number of accesses to all its elements equal to $m$, and let $T'$ be the direct subtree of $T$. Then, $m(T')$ is $C \cdot (\frac{m}{\operatorname{D}(m) + 1})$ with probability at least $1 - O(\frac{\operatorname{D}(m) + 1}{m})$ and $C \leq \frac{5}{4}$.
\end{lemma}
\begin{proof}
Let $m_0$ be the total number of requests to $T$ after its last rebuild. Then, $m \leq \frac{5}{4} \times m_0$ and $c \leq \frac{1}{4} \times m_0$ accesses were made into $T$ since it was rebuilt for the last time. When $T$ was rebuilt for the last time, $T'$ had no more than $ \frac{m_0}{\operatorname{D}(m_0) + 1}$ accesses. Suppose that $X$ additional accesses were made in $T'$. 

From here on let us treat these accesses as separate elements, transforming probability mass function $p$ into probability density function $\mu$, mapping integer $x$ with $ac$ accesses into $ac$ elements drawn uniformly from range $[x, x + 1)$.

\emph{This transformation is called \textbf{$p$-extension} and we denote it as $\mu = \operatorname{Extension}(p)$.} After that $T'$ is a $\mu'[a, b]$-random GSAT, whose elements are drawn in accordance with density function $$\mu'(x) = \frac{\mu(x)}{\int_c^d{\mu(y) dy}}.$$

From this point of view, no more than $\frac{m_0}{\operatorname{D}(m_0) + 1}$ elements were stored in $T'$ after the last rebuild and $X$ additional items were stored in $T'$, so we are in the same configuration as in~\cite[Lemma 5]{DynamicIST}, thus, we can repeat it: $$E(X) = \frac{m_0}{\operatorname{D}(m_0) + 1} \cdot \frac{c}{m_0 + 1} \leq \frac{1}{4} \times \frac{m_0}{\operatorname{D}(m_0) + 1},$$ $$\operatorname{VAR}(X) = \sigma^2 \leq E(X) \cdot \Big(1 + \frac{c}{m_0 + 1}\Big) \leq \frac{1}{2} \times \frac{m_0}{\operatorname{D}(m_0) + 1}.$$

Using Chebyshev's inequality $Pr(|X - E(X)| \geq t) \leq \frac{\sigma^2}{t^2}$ with $t = \frac{m_0}{\operatorname{D}(m_0) + 1}$ we have $$Pr\Big(X \geq E(X) + \frac{m_0}{\operatorname{D}(m_0) + 1}\Big) \leq \frac{1}{2} \times \frac{m_0}{\operatorname{D}(m_0) + 1} \cdot \Big(\frac{\operatorname{D}(m_0) + 1}{m_0}\Big)^2 = \frac{1}{2} \times \frac{\operatorname{D}(m_0) + 1}{m_0},$$
so, the probability that more than $\frac{5}{4} \times \frac{m}{\operatorname{D}(m) + 1}$ accesses were made into the tree $T'$ is $\leq \frac{1}{2} \times \frac{\operatorname{D}(m) + 1}{m}$, so, $m(T') = C \cdot \frac{m}{\operatorname{D}(m) + 1}$ with probability $1 - O(\frac{\operatorname{D}(m) + 1}{m})$.
\end{proof}

Lemma~\ref{lem:prob-size} illustrates the self-organizing feature of GSATs and now we are ready to prove the expected total cost of executed operations.

\begin{lemma}
\label{lem:expected-depth}
Let $p$ be a probability mass function with finite support $[a, b]$. Then, the expected total cost of processing a sequence of $m$ $p$-random insertions, $p$-random deletions, and $p$-random gets to an initially empty GSAT (not counting the time for the search in the node) is $O(m \cdot d)$, where $d$ is the depth of the ideal GSAT. Thus, the expected amortized cost of \texttt{insert}, \texttt{delete}, or \texttt{get} is $O(d)$. 
\end{lemma}
\begin{proof}
Let $f(m)$ be the expected
number of traversed nodes by the $m$-th operation. Then, for some constant $C \leq \frac{5}{4}$:
$$f(m) \le 1 + f\Big(C \cdot \Big( \frac{m}{\operatorname{D}(m) + 1} \Big)\Big) + O(\log m) \times O\Big(\frac{\operatorname{D}(m) + 1}{m}\Big).$$

This can be seen as follows: if the operation goes into a subtree of size $C \cdot \frac{m}{\operatorname{D}(m) + 1}$, then, the operation traverses $1 + f(C \cdot \frac{m}{\operatorname{D}(m) + 1})$ nodes. If it does not, then it traverses at most $O(\log m)$ nodes by Lemma~\ref{lem:gsat-total-depth}. The probability of the latter event is $O(\frac{\operatorname{D}(m) + 1}{m})$ by Lemma~\ref{lem:prob-size}, therefore, since $\operatorname{D}(m)$ is sqrt-bounded: $$O(\log m) \times O\Big(\frac{\operatorname{D}(m) + 1}{m}\Big) \leq O(\log m) \times O\Big( \frac{2}{\sqrt{m}} \Big) = O\Big( \frac{2 \cdot \log m}{\sqrt{m}} \Big) = O(1).$$ 

A subtree of ideal GSAT has no more than $\frac{m}{\operatorname{D}(m) + 1}$ accesses, however, in the recurrence for $f(m)$ on the next step we have $C \cdot \frac{m}{\operatorname{D}(m) + 1}$ accesses. But we show that it takes just two steps to reach $\frac{m}{\operatorname{D}(m) + 1}$ since $C \leq \frac{5}{4}$: $$ C \cdot \frac{m}{\operatorname{D}(m) + 1} \longrightarrow \frac{C^2 \cdot m}{(\operatorname{D}(m) + 1) \cdot (\operatorname{D}(\ldots) + 1)} \leq \frac{C^2}{2} \times \frac{m}{\operatorname{D}(m) + 1} < \frac{m}{\operatorname{D}(m) + 1}.$$

Thus, $f(m) = O(d)$.
\end{proof}

Using this Lemma together with Lemma~\ref{lem:gsat-nice-depth} we get the following result.

\begin{theorem}
\label{lem:depth-element}
An expected depth of key $x$ with $ac$ requests in a tree with $m$ requests is $O(\log \frac{\log m}{\log ac})$ in SAIT, and $O(\log_B \frac{m}{ac})$ in SABT.
\end{theorem}
\begin{proof}
First, we consider SAIT. On the depth $d$ the number of requests to a subtree does not exceed $m^{\frac{1}{2^d}}$. Comparing it with $ac$ we get that the depth $d$ of key $x$ does not exceed $O(\log \frac{\log m}{\log ac})$.

The same happens with SABT. On the depth $d$ the number of requests to a subtree does not exceed $\frac{m}{B^d}$. Comparing it with $ac$ we get the depth $d$ of key $x$ does not exceed $O(\log_B \frac{m}{ac})$.
\end{proof}

\section{Self-Adjusting IST (SAIT)}
\label{sec:sait}

\begin{theorem}
\label{th:sait-construction}
An ideal SAIT requires $O(m)$ time to build and uses $O(m^{\frac{\alpha}{2}} \times n + m^{\alpha})$ memory.
\end{theorem}
\begin{proof}
For SAIT, we can build all nodes with representatives in $O(m)$ as shown in Theorem~\ref{th:gsat-construction}. However, we need to add the construction time for $ID$ array in each node. Let us denote $P(m)$ time to build $ID$ arrays in the whole SAIT, then we have $P(m) = O(m^\alpha) + (\sqrt{m} + 1) \times P(\frac{m}{\sqrt{m} + 1})$. Let $\widetilde{P}(m) = \frac{P(m)}{m}$, so $m \cdot \widetilde{P}(m) = P(m)$ and $\widetilde{P}(m) = O(m^{\alpha - 1}) + \widetilde{P}(\frac{m}{\sqrt{m} + 1}) \leq O(m^{\alpha - 1}) + \widetilde{P}(\sqrt{m})$, thus, if $m = 2^{2^r}$: $$\widetilde{P}(m) = O\Big( \sum\limits_{i = 0}^r (m^{\frac{1}{2^i}})^{\alpha - 1} \Big) = O\Big( \sum\limits_{i = 0}^r (2^{2^i})^{\alpha - 1} \Big) = O(1),$$ since $\alpha < 1$. Therefore, $P(m) = O(m)$, which gives us $O(m)$ time to build an ideal SAIT.

Now, let us count the memory consumption. The root node of SAIT requires $O(m^{\alpha})$ space to store representatives of size $\sqrt{m}$ and $ID$ array of size $m^{\alpha}$. Also, the total number of nodes in SAIT does not exceed $n$ since the tree is internal. For all other subtrees $m(T_i) \leq \sqrt{m}$. Thus, $ID$ for each of these trees has a size no more than $m^{\frac{\alpha}{2}}$, and, therefore, the memory required for all of the nodes, except for the root, does not exceed $O(m^{\frac{\alpha}{2}} \times n)$. Summing everything up we get the required bound.
\end{proof}

\begin{theorem}
\label{th:sait-update-mem}
After \texttt{insert}, \texttt{delete}, and \texttt{get} operations, the SAIT for the set of $n$ elements with $m$ total number of accesses consumes $O(m^{\alpha} \times n)$ space.
\end{theorem}
\begin{proof}
If after any operation above no subtrees were rebuilt, then we spend $O(1)$ in case of \texttt{insert} operation since the new node consumes only a constant memory. Otherwise, some subtree $T'$ was rebuilt. After rebuilding $T'$, memory has changed only for $T'$ and its subtrees. If $T'$ had $n'$ elements, after rebuilding it can not spend more than $O(m^{\alpha} \times n')$, since for each element we can not spend more memory than for $ID$ array, whose maximum size is $O(m^\alpha)$, thus, we get the required bound.
\end{proof}

\subsection{Expected time analysis of the search function}
\label{subsubsec:search}



The standard search operation in a node of GSAT uses binary search. However, in SAIT, similar to~\cite{DynamicIST}, we can speed up $\operatorname{S}(T, key)$ for a large class of density functions, obtained after applying $p$-extension.

\begin{definition}[\cite{DynamicIST}]
\label{def:smooth}
A density $\mu$ is \texttt{smooth} for a parameter $\alpha$, $\frac{1}{2} \leq \alpha < 1$, if there are constants $a$, $b$ and $d$ such that $\mu(x) = 0$ for $x < a$ and $x > b$ and such that for all $c_1$, $c_2$, $c_3$, $a \le c_1 < c_2 < c_3 \le b$, and all integers $n$ and $m$ with $m = \lceil n^{\alpha} \rceil$, 
    $$\int_{c_2 - \frac{c_3 - c_1}{m}}^{c_2}\mu[c_1, c_3](x)\,dx \le d \times n^{-\frac{1}{2}},$$
where $\mu[c_1, c_3](x) = 0$ for $x < c_1$ or $x > c_3$ and $\mu[c_1, c_3](x) = \frac{\mu(x)}{p}$ for $c_1 \le x \le c_3$ where $p = \int_{c_1}^{c_3}\mu(x)dx$.
\end{definition}

An array $ID$, used for SAIT, is initialized and used in the same way as in~\cite{DynamicIST}. Then, the proof of the next lemma can be repeated.

\begin{lemma}
\label{lem:root-search}
Let $T$ be a $p[a, b]$-random SAIT, $\mu = \operatorname{Extension}(p)$ and $\mu$ is a smooth density with parameter $\alpha$. Then, the expected search time in the root array is $O(1)$, i.e., $Time(\operatorname{S}(T, key)) = O(1)$.
\end{lemma}



Now, taking the depth bound from Theorem~\ref{lem:depth-element} and the expected search time in a node from Lemma~\ref{lem:root-search} we get the following final result.

\begin{theorem}
\label{th:sait-nice-depth}
The expected search time of a key with the number of accesses $ac$ in a \\ $p[a, b]$-random SAIT, where $\mu = \operatorname{Extension}(p)$ is a smooth density for a parameter $\alpha$, is $O(\log{\frac{\log{m}}{\log{ac}}})$.
\end{theorem}



\section{Self-Adjusting B-Tree (SABT)}
\label{sec:sabt}

\begin{theorem}
\label{th:sabt-construction}
An ideal SABT has $O(\min(n\log n, m))$ construction time, has $O(n)$ memory, and has depth $O(\log_B(m))$.
\end{theorem}
\begin{proof}
We have $O(n \log n)$ bound for the construction time, since there are only $n$ elements in the tree, each one is selected as a representative exactly once using the binary search with $O(\log n)$ complexity, so we get $O(\min(n\log n, m))$ time to build and $O(n)$ memory. For the depth we get $d(m) \leq 1 + d(\frac{m}{B})$, so $d(m) = O(\log_B(m))$.
\end{proof}

\begin{theorem}
\label{th:sabt-update-mem}
After \texttt{insert}, \texttt{delete}, and \texttt{get} operations, the SABT for the set of $n$ elements consumes $O(n)$ space.
\end{theorem}
\begin{proof}
SABT is an internal tree, thus, if after an operation no tree was rebuilt, we spend $O(1)$ in case of \texttt{insert} operation, and if some subtree $T'$ with size $n'$ was rebuilt, it consumes $O(n')$ space.
\end{proof}

As for the SABT, using Lemma~\ref{lem:depth-element}, we can get the following result when the search function in the node takes $O(\log B)$:
\begin{theorem}
The expected search time of a key with the number of accesses $ac$ in a SABT with a parameter $B$ is $O(\log_2{\frac{m}{ac}})$.
\end{theorem}

\section{Self-Adjusting Log Tree (SALT)}
\label{sec:salt}

Instead of taking $D(m)$ as $\sqrt{m}$ or $B$, we can parameterize with another function, for example, $\log m$. Thus, let $D(m) = \log m$ and the search function is the binary search, we name this tree a Self-Adjusting Log Tree (SALT).

\begin{theorem}
\label{th:salt-construction}
An ideal SALT requires $O(\min(n\log n, m))$ time to build, uses $O(n)$ memory, and has depth $O(\frac{\log m}{\log \log m})$.
\end{theorem}
The proof can be found in Appendix~\ref{app:salt-proof}.

Repeating the proof of Theorem~\ref{th:sabt-update-mem}, we get:
\begin{theorem}
\label{th:salt-update-mem}
After \texttt{insert}, \texttt{delete}, and \texttt{get} operations, the SALT for the set of $n$ elements consumes $O(n)$ space.
\end{theorem}

\section{Range queries}
\label{sec:range-queries}

In this section, we extend our GSAT to support range queries. The main advantage of GSAT range queries, in comparison with other self-adjusting data structures, is that if the query affects keys $[x_{i_1}, x_{i_2}, \ldots x_{i_z}]$, the number of accesses to all of them increases by one ($j = 1,2,\ldots,z: ac_{i_j}' = ac_{i_j} + 1$), so, the future requests to them become faster.

For the description of the range queries and the complexity analysis, we introduce the following notations:
\begin{itemize}
    \item $\operatorname{get}(T, a, b)$ as \texttt{get(a, b)} applied to the tree $T$;
    \item $\operatorname{calculate}(T, a, b)$ as \texttt{calculate(a, b)} applied to the tree $T$;
    \item $\operatorname{update}(T, a, b, c)$ as \texttt{update(a, b, c)} applied to the tree $T$;
    \item $\operatorname{CollectNonMarked}(T)$ as ordered keys $[x_{j_1}, x_{j_2}, \ldots, x_{j_z}]$ that are non-marked and belong to the tree $T$;
    \item $lb(T)$ / $rb(T)$ as a left / right bound of the segment on which tree $T$ was built;
    \item $x_{\geq a}$ as the first key in the tree whose value is not less than $a$;
    \item $x_{\leq b}$ as the first key in the tree whose value is not greater than $b$;
    \item $d(x)$ as the depth of key $x$;
    \item $y_i$ as a value of the key $x_i$.
\end{itemize}

We start with the description of \texttt{get(a, b)} query that retrieves present keys from the segment $[a, b]$ and for which we do not have to modify the GSAT node in any way. 

For $\operatorname{get}(T, a, b)$, at the start, we have to find the subtree $T_i$ such that $lb(T_i) < a \leq rb(T_i)$, using the search function $\operatorname{S}(T, a)$. If $b < rb(T_i)$ then we return $\operatorname{get}(T_i, a, b)$. Otherwise, we perform search $\operatorname{S}(T, b)$ to find the subtree $T_j$ such that $lb(T_j) \leq b < rb(T_j)$. Then, the answer to \texttt{get(a, b)} query is: $$[\operatorname{get}(T_i, a, rb(T_i)), REP[i], \operatorname{CollectNonMarked}(T_{i+1}),REP[i + 1],\ldots, REP[j - 1], \operatorname{get}(T_j, lb(T_j), b)],$$ where we include only non-marked keys $REP[\cdot]$.

Moreover, we do not need to find the right bound for $T_i$, while doing $\operatorname{get}(T_i, a, rb(T_i))$, since it is always the rightmost bound as well as the left bound for $\operatorname{get}(T_j, lb(T_j), b)$ since it is always the leftmost bound. Also, for each added $REP[\cdot]$ we increment the number of accesses to it as well as increment counters for each traversed node and perform rebuilding as for \texttt{insert}, \texttt{delete}, or \texttt{get} operations.

\begin{theorem}
\label{th:get-range-query}
The time complexity of \texttt{get(a, b)} range query is $O(\ell + \sum\limits_{i=0}^{d} \operatorname{S}(T^{(i)}, key))$, where there are $\ell$ keys in the tree, which belongs to the segment $[a, b]$, and $d = \max(d(x_{\geq a}), d(x_{\leq b}))$.
\end{theorem}
\begin{proof}
While $lb(T') < a < b < rb(T')$ we make one search at the node. Otherwise, the query is split into $\operatorname{get}(T_i, a, rb(T_i))$ and $\operatorname{get}(T_j, lb(T_j), b)$. Then, we make at most one search at the node for: 1)~$\operatorname{get}(T_i, a, rb(T_i))$ to find the next left bound, until $x_{\geq a}$ found~--- after that, there is no subtree to visit; 2)~$\operatorname{get}(T_j, lb(T_j), b)$ to find the next right bound, until $x_{\leq b}$ found. Each $\operatorname{CollectNonMarked}$ visits the keys from the segment $[a, b]$. Thus, we get the required complexity.
\end{proof}

Next, we explain the implementation of \texttt{calculate(a, b)} and \texttt{update(a, b, c)} range queries on the high level.
$\operatorname{calculate}$ performs an operation $\odot$ on all the values with the keys in the range, for example, \textrm{sum}, and $\operatorname{update}$ apply $(\star c)$ to all the values with the keys in the range, for example, \textrm{set $c$}.

For these operations, we extend each GSAT node with four segment trees~\cite{SegmentTree}:
\begin{enumerate}
    \item $ST_m$ to maintain the number of requests, built on an array \\ $[m(T_1), AC[1], m(T_2), AC[2], \ldots, AC[|REP|], m(T_{|REP| + 1})]$;
    \item $ST_f$ to compute the result of applied functions $\odot$ and $\star$, built on $[\odot(T_1), y_{i_1}, \odot(T_2), y_{i_2}, \ldots, y_{i_k}, \odot(T_{|REP| + 1})]$;
    \item $ST_{pm}$ to propagate the number of requests to the subtrees $T_1, T_2, \ldots, T_{|REP| + 1}$;
    \item $ST_{pf}$ to propagate $\star$ operation to the subtrees $T_1, T_2, \ldots, T_{|REP| + 1}$.
\end{enumerate}
All segment trees consume $O(|REP|)$ memory per node, thus, they do not affect the memory and the construction time complexities of our data structures.

To $\operatorname{calculate}(T, a, b)$ and $\operatorname{update}(T, a, b, c)$, we find $T_i$ and $T_j$ in the same way as for $\operatorname{get}(T, a, b)$. However, while going down to the subtree, we have to also propagate the number of requests as well as apply $\star$ from previous operations~--- for that, we use two segment trees $ST_{pm}$ and $ST_{pf}$.

Therefore, for $\operatorname{calculate}(T, a, b)$ query: 
\begin{enumerate}
    \item the final result is: $\operatorname{calculate}(T_i, a, rb(T_i)) \, \odot \, ST_f[y_i, T_{i + 1}, y_{i + 1}, \ldots, T_{j - 1}, y_{j - 1}] \, \odot \, \operatorname{calculate}(T_j, lb(T_j), b)$;
    \item we have to increment the number of requests through $ST_{pm}[T_{i + 1}, \ldots, T_{j - 1}]$.
\end{enumerate}

And for $\operatorname{update}(T, a, b, c)$ query we perform the following:
\begin{enumerate}
    \item call $ST_f[y_i, T_{i + 1}, y_{i + 1}, \ldots, T_{j - 1}, y_{j - 1}]$ and $ST_{pf}[T_{i + 1}, \ldots, T_{j - 1}]$ to accumulate operation $(\star c)$;
    \item call $\operatorname{update}(T_i, a, rb(T_i), c)$ and $\operatorname{update}(T_j, lb(T_j), b, c)$;
    \item we have to increment the number of requests through $ST_{pm}[T_{i + 1}, \ldots, T_{j - 1}]$. 
\end{enumerate}

To combine $\odot$ and $\star$ operations our segment trees use the lazy propagation technique~\cite{LazySegmentTree}. Obviously, we increment counters for each traversed node during \texttt{calculate(a, b)} or \texttt{update(a, b, c)} queries and perform rebuilding as for \texttt{insert}, \texttt{delete}, or \texttt{get} operations, not forgetting to propagate the number of requests and the calculated function onto the values.

\begin{theorem}
\label{th:calculate-update-range queries}
The time complexity of \texttt{calculate(a, b)} or \texttt{update(a, b, c)} range queries is $O(\log n \cdot  \max(d(x_{\geq a}), d(x_{\leq b})))$.
\end{theorem}
\begin{proof}
While traversing the nodes, at each node we make $O(1)$ requests to our segment trees to compute either applied operations or operations to propagate. On each request, we spend $O(\log n)$, because for each node $|REP| \leq n$. We stop the traversal when we reach $d(x_{\geq a})$ or $d(x_{\leq b}))$ for the same reasons as for \texttt{get(a, b)} query in Theorem~\ref{th:get-range-query}.
\end{proof}

Now, we can state that such range queries complicate all the previous operations because we additionally compute the number of requests for propagation that is done in $O(\log n)$. Moreover, for \texttt{insert} and \texttt{delete} operations we have to update all segment trees on the traversed path that covers the argument key.

Storing segment trees helps us to speed up operations on large nodes, but if $\operatorname{D}(m) = O(1)$, i.e., for B-Tree, we can get rid of them and spend the constant time in each node, achieving $O(\max(d(x_{\geq a}), d(x_{\leq b})))$ per operation.

\section{Concurrent lock-free transformation}
\label{sec:concurrent}

To make GSAT concurrent, we use a collaborative rebuild approach from~\cite{CIST}. The main difference from the paper is that we now store the number of requests instead of the sizes. Thus, in the function \texttt{markAndCount} from \cite{CIST} we count the number of accesses made to subtrees and then choose representatives on the first level. Next, we use the collaborative rebuild, changing the initial builder to our GSAT builder that uses binary searches on the prefix sums of accesses to split subtrees. This extension supports \texttt{insert}, \texttt{delete}, and \texttt{get} operations, but not range queries.

\section{Experiments and results}
\label{sec:experiments}

In this section, we compare described GSAT implementations with their classic versions and, additionally, with Splay Tree~\cite{splay-tree}. 
These workloads were executed on a system with Intel Xeon Gold 6240R CPU @ 2.40GHz, RAM 256Gb.

\subsection{Workloads}

Experiments were carried out on skewed \texttt{x/y} workloads and heavy-tailed zipf distribution workload, parameterized with $1$. We described the first ones while the last one is well-known.


For \texttt{x/y} workload, let us denote $S$ as the whole set of keys and $S_y$ as a subset of $S$, having size $y \cdot |S|$~--- the key is taken with probability $y\%$. In that workload, $x\%$ of operations choose an argument from $S_y$ while the rest of operations choose an argument from $S \setminus S_y$. In other words, the more difference between $x$ and $y$ is, the more the workload is skewed.

For experiments we consider only map operations: \texttt{get(key)}, \texttt{insert(key, value)}, and \texttt{delete(key)}. We use two settings with different amounts of update operations: 
\begin{enumerate}
    \item $0\%$, i.e., only \texttt{get} operations. We call such workloads as \emph{read-only}.
    \item $20\%$, i.e., $10\%$ of \texttt{insert} operations, $10\%$ of \texttt{delete} operations, and $80\%$ of \texttt{get} operations. We call such workloads as \emph{mixed}.
\end{enumerate}
The operations are chosen uniformly with respect to the proportions.

We consider workloads with different skewness: \texttt{100/100}, i.e., uniform; \texttt{70/30}; \texttt{80/20}; \texttt{90/10}; \texttt{95/05}; and \texttt{99/01}.
While for the size of $S$ we use three settings: $10^5$, $10^6$, and $10^7$.


\subsection{Tables with experiments}

Each cell presents the throughput (or the relative throughput) of the corresponding data structure on the corresponding workload. Throughput equals the total number of operations divided by working time. We run each experiment for $20$ seconds. To remove any fluctuations we run each experiment $5$ times, and the average value appears in the table.

We choose Splay Tree as the baseline and the throughput of other data structures is calculated relative to it.

B-Tree and IST correspond to the original B-Tree with $B = 8$ (i.e., each node stores $8-16$ keys) and IST implementations. SAIT, SABT with $B = 16$, and SALT are our self-adjusting data structures which we present in the paper. Finally, we also compare SA2T which is the self-adjusting binary tree and is implemented as SABT with $B = 2$.

We start with read-only experiments, i.e., only \texttt{get} operations: Tables~\ref{table:get-10^5},~\ref{table:get-10^6}, and~\ref{table:get-10^7}.

\begin{table}[!h]
\caption{read-only workloads, key set $|S| = 10^5$}
\label{table:get-10^5}
\centering
\resizebox{\linewidth}{!}{%
\begin{tabular}{|c | c | c | c | c | c | c | c|}
\hline
 & uniform & 70/30 & 80/20 & 90/10 & 95/05 & 99/01 & zipf-1\\
\hline
Splay Tree & $2.6 \cdot {10}^{6}$ & $2.4 \cdot {10}^{6}$ & $2.7 \cdot {10}^{6}$ & $3.4 \cdot {10}^{6}$ & $4.1 \cdot {10}^{6}$ & $6.2 \cdot {10}^{6}$ & $5.7 \cdot {10}^{7}$ \\
\hline
SA2T & x$1.43$ & x$1.48$ & x$1.39$ & x$1.4$ & x$1.44$ & \cellcolor{blue!20}x$1.41$ & \cellcolor{blue!20}x$0.86$ \\
\hline
B-Tree & x$2.64$ & x$2.43$ & x$2.28$ & x$1.98$ & x$1.74$ & x$1.35$ & x$0.48$ \\
\hline
SABT & \cellcolor{blue!20}x$2.52$ & \cellcolor{blue!20}x$2.32$ & x$2.15$ & \cellcolor{blue!20}x$1.93$ & \cellcolor{blue!20}x$1.84$ & \cellcolor{blue!20}x$1.79$ & \cellcolor{blue!20}x$0.78$ \\
\hline
IST & x$1.93$ & x$1.89$ & x$1.75$ & x$1.54$ & x$1.34$ & x$1.08$ & x$0.23$ \\
\hline
SAIT & \cellcolor{blue!20}x$3.4$ & \cellcolor{blue!20}x$2.99$ & \cellcolor{blue!20}x$2.76$ & \cellcolor{blue!20}x$2.57$ & \cellcolor{blue!20}x$3.83$ & \cellcolor{blue!20}x$3.33$ & \cellcolor{blue!20}x$0.5$ \\
\hline
SALT & x$2.14$ & x$1.85$ & x$1.75$ & x$1.8$ & \cellcolor{blue!20}x$1.8$ & \cellcolor{blue!20}x$1.68$ & \cellcolor{blue!20}x$0.74$ \\
\hline
\end{tabular}
}
\end{table}

\begin{table}[!h]
\caption{read-only workloads, key set $|S| = 10^6$}
\label{table:get-10^6}
\centering
\resizebox{\linewidth}{!}{%
\begin{tabular}{|c | c | c | c | c | c | c | c|}
\hline
 & uniform & 70/30 & 80/20 & 90/10 & 95/05 & 99/01 & zipf-1\\
\hline
Splay Tree & $1 \cdot {10}^{6}$ & $9.7 \cdot {10}^{5}$ & $1 \cdot {10}^{6}$ & $1.4 \cdot {10}^{6}$ & $2 \cdot {10}^{6}$ & $3.7 \cdot {10}^{6}$ & $5.5 \cdot {10}^{7}$ \\
\hline
SA2T & x$1.18$ & x$1.16$ & x$1.18$ & x$1.1$ & x$1.1$ & \cellcolor{blue!20}x$1.21$ & \cellcolor{blue!20}x$0.85$ \\
\hline
B-Tree & x$2.61$ & x$2.2$ & x$2.22$ & x$1.88$ & x$1.61$ & x$1.27$ & x$0.39$ \\
\hline
SABT & x$2.44$ & x$2.02$ & \cellcolor{blue!20}x$2.21$ & \cellcolor{blue!20}x$2.0$ & \cellcolor{blue!20}x$2.22$ & \cellcolor{blue!20}x$1.84$ & \cellcolor{blue!20}x$0.79$ \\
\hline
IST & x$1.43$ & x$1.32$ & x$1.42$ & x$1.18$ & x$1.01$ & x$0.83$ & x$0.24$ \\
\hline
SAIT & x$1.88$ & x$1.7$ & x$1.95$ & \cellcolor{blue!20}x$2.13$ & \cellcolor{blue!20}x$2.1$ & \cellcolor{blue!20}x$2.16$ & \cellcolor{blue!20}x$0.49$ \\
\hline
SALT & x$1.75$ & x$1.54$ & x$1.56$ & x$1.76$ & \cellcolor{blue!20}x$1.77$ & \cellcolor{blue!20}x$1.73$ & \cellcolor{blue!20}x$0.72$ \\
\hline
\end{tabular}
}
\end{table}

\begin{table}[H]
\caption{read-only workloads, key set $|S| = 10^7$}
\label{table:get-10^7}
\centering
\resizebox{\linewidth}{!}{%
\begin{tabular}{|c | c | c | c | c | c | c | c|}
\hline
 & uniform & 70/30 & 80/20 & 90/10 & 95/05 & 99/01 & zipf-1\\
\hline
Splay Tree & $4.7 \cdot {10}^{5}$ & $4.3 \cdot {10}^{5}$ & $4.6 \cdot {10}^{5}$ & $5.7 \cdot {10}^{5}$ & $7.3 \cdot {10}^{5}$ & $1.7 \cdot {10}^{6}$ & $5.6 \cdot {10}^{7}$ \\
\hline
SA2T & x$1.05$ & x$1.09$ & x$1.15$ & x$0.93$ & x$0.75$ & x$0.49$ & \cellcolor{blue!20}x$0.55$ \\
\hline
B-Tree & x$2.92$ & x$2.84$ & x$2.71$ & x$2.21$ & x$1.84$ & x$0.92$ & x$0.34$ \\
\hline
SABT & x$2.16$ & x$2.1$ & x$2.17$ & x$1.89$ & \cellcolor{blue!20}x$1.77$ & \cellcolor{blue!20}x$1.03$ & \cellcolor{blue!20}x$0.6$ \\
\hline
IST & x$1.66$ & x$1.87$ & x$1.71$ & x$1.45$ & x$1.08$ & x$0.52$ & x$0.18$ \\
\hline
SAIT & x$1.74$ & x$1.6$ & x$1.52$ & x$1.52$ & x$1.32$ & \cellcolor{blue!20}x$1.4$ & \cellcolor{blue!20}x$0.34$ \\
\hline
SALT & x$1.66$ & x$1.42$ & x$1.4$ & x$1.44$ & x$1.18$ & x$0.73$ & \cellcolor{blue!20}x$0.48$ \\
\hline
\end{tabular}
}
\end{table}

Colored cells indicate GSAT values that are very close (lose no more than $5\%$) or noticeably better than those of non-self-adjusting trees (note that Splay Tree is the self-adjusting tree). 

As we can see, the more skewed the access distribution becomes, the better GSATs work. In addition, the smaller the set of keys, the more requests occurred to the keys and the faster our GSATS work~--- this can be determined by the number of colored cells as the size of the set of keys increases. SAIT is superior to other trees, because, on average, it takes constant time to search in the node due to the $ID$ array, while its depth is much less. This is expected since our workloads are \texttt{smooth} as shown in Lemma~\ref{lem:bounded-distr}.

GSATs surpass Splay Tree on every \texttt{x/y} workload, however, on zipf-1 workload, they are inferior in performance. In heavy-tailed workloads, Splay Tree works only with keys near the root which is also the case for our data structures, but they have an additional overhead spent on tree rebuilds.

Then, we consider mixed queries with $20\%$ of update operations, i.e., $10\%$ of \texttt{insert} operations, $10\%$ of \texttt{delete} operations, and $80\%$ of \texttt{get} operations. This workload is typically used to mimic loads on databases~\cite{aksenov2017concurrency}. The results are presented on Tables~\ref{table:mixed-10^5},~\ref{table:mixed-10^6}, and~\ref{table:mixed-10^7}.

\begin{table}[!h]
\caption{mixed workloads, key set $|S| = 10^5$}
\label{table:mixed-10^5}
\centering
\resizebox{\linewidth}{!}{%
\begin{tabular}{|c | c | c | c | c | c | c | c|}
\hline
 & uniform & 70/30 & 80/20 & 90/10 & 95/05 & 99/01 & zipf-1\\
\hline
Splay Tree & $2.5 \cdot {10}^{6}$ & $2.2 \cdot {10}^{6}$ & $2.4 \cdot {10}^{6}$ & $2.8 \cdot {10}^{6}$ & $3.2 \cdot {10}^{6}$ & $4.3 \cdot {10}^{6}$ & $4.4 \cdot {10}^{7}$ \\
\hline
SA2T & x$1.17$ & x$1.15$ & x$1.07$ & x$1.04$ & x$0.95$ & x$0.72$ & x$0.1$ \\
\hline
B-Tree & x$2.61$ & x$2.47$ & x$2.33$ & x$2.12$ & x$1.97$ & x$1.62$ & x$0.53$ \\
\hline
SABT & x$2.26$ & x$2.09$ & x$2.0$ & x$1.84$ & x$1.76$ & x$1.49$ & \cellcolor{blue!20}x$0.59$ \\
\hline
IST & x$1.65$ & x$1.58$ & x$1.47$ & x$1.35$ & x$1.27$ & x$1.13$ & x$0.21$ \\
\hline
SAIT & x$1.93$ & x$1.78$ & x$1.73$ & x$1.67$ & x$1.57$ & x$1.39$ & x$0.26$ \\
\hline
SALT & x$1.56$ & x$1.51$ & x$1.44$ & x$1.4$ & x$1.3$ & x$1.14$ & x$0.26$ \\
\hline
\end{tabular}
}
\end{table}

\begin{table}[!h]
\caption{mixed workloads, key set $|S| = 10^6$}
\label{table:mixed-10^6}
\centering
\resizebox{\linewidth}{!}{%
\begin{tabular}{|c | c | c | c | c | c | c | c|}
\hline
 & uniform & 70/30 & 80/20 & 90/10 & 95/05 & 99/01 & zipf-1\\
\hline
Splay Tree & $1 \cdot {10}^{6}$ & $9.4 \cdot {10}^{5}$ & $9.1 \cdot {10}^{5}$ & $1.1 \cdot {10}^{6}$ & $1.4 \cdot {10}^{6}$ & $2.5 \cdot {10}^{6}$ & $4.4 \cdot {10}^{7}$ \\
\hline
SA2T & x$1.03$ & x$0.99$ & x$1.04$ & x$0.84$ & x$0.7$ & x$0.48$ & x$0.07$ \\
\hline
B-Tree & x$2.63$ & x$2.02$ & x$2.31$ & x$2.09$ & x$1.89$ & x$1.52$ & x$0.44$ \\
\hline
SABT & x$2.19$ & x$1.91$ & x$2.06$ & x$1.96$ & \cellcolor{blue!20}x$1.86$ & \cellcolor{blue!20}x$1.45$ & x$0.33$ \\
\hline
IST & x$1.46$ & x$1.24$ & x$1.3$ & x$1.16$ & x$1.01$ & x$0.99$ & x$0.21$ \\
\hline
SAIT & x$1.48$ & x$1.33$ & x$1.39$ & x$1.23$ & x$1.17$ & x$1.06$ & x$0.19$ \\
\hline
SALT & x$1.36$ & x$1.3$ & x$1.35$ & x$1.18$ & x$1.04$ & x$0.83$ & x$0.15$ \\
\hline
\end{tabular}
}
\end{table}

\begin{table}[H]
\caption{mixed workloads, key set $|S| = 10^7$}
\label{table:mixed-10^7}
\centering
\resizebox{\linewidth}{!}{%
\begin{tabular}{|c | c | c | c | c | c | c | c|}
\hline
 & uniform & 70/30 & 80/20 & 90/10 & 95/05 & 99/01 & zipf-1\\
\hline
Splay Tree & $4.3 \cdot {10}^{5}$ & $4 \cdot {10}^{5}$ & $4.2 \cdot {10}^{5}$ & $4.9 \cdot {10}^{5}$ & $6 \cdot {10}^{5}$ & $9.6 \cdot {10}^{5}$ & $4.4 \cdot {10}^{7}$ \\
\hline
SA2T & x$1.05$ & x$1.13$ & x$1.17$ & x$1.02$ & x$0.79$ & x$0.56$ & x$0.03$ \\
\hline
B-Tree & x$3.09$ & x$3.0$ & x$2.87$ & x$2.41$ & x$1.95$ & x$1.48$ & x$0.4$ \\
\hline
SABT & x$2.26$ & x$2.35$ & x$2.34$ & x$2.02$ & x$1.62$ & x$1.25$ & x$0.28$ \\
\hline
IST & x$1.73$ & x$1.83$ & x$1.71$ & x$1.58$ & x$1.17$ & x$0.79$ & x$0.12$ \\
\hline
SAIT & x$1.94$ & x$1.57$ & x$1.55$ & x$1.51$ & x$1.34$ & x$0.89$ & x$0.11$ \\
\hline
SALT & x$1.69$ & x$1.45$ & x$1.41$ & x$1.49$ & x$1.26$ & x$0.81$ & x$0.09$ \\
\hline
\end{tabular}
}
\end{table}

On mixed workloads GSATs continue to outperform Splay Tree on every \texttt{x/y} workload, however, our GSATs, typically, work worse than non-self-adjusting trees and even do not give an advantage on zipf-1 workload.

It happens due to \texttt{delete} operations: after rebuilding a tree we physically remove all marked (deleted) keys and they lose their accesses, therefore, the following operations to them will be slow. This is more serious for GSATs than for Splay Tree, since in Splay Tree the newly added key immediately becomes the root node, and the next operations to it will be fast.

To overcome this problem, we launched the same mixed workloads, but now our GSATs do not physically delete keys~--- all keys remain physically present even after the rebuilding, remembering the number of previous accesses. The results are shown on Tables~\ref{table:lazy-delete-mixed-10^5},~\ref{table:lazy-delete-mixed-10^6}, and~\ref{table:lazy-delete-mixed-10^7}.

\begin{table}[!h]
\caption{lazy-delete mixed workloads, key set $|S| = 10^5$}
\label{table:lazy-delete-mixed-10^5}
\centering
\resizebox{\linewidth}{!}{%
\begin{tabular}{|c | c | c | c | c | c | c | c|}
\hline
 & uniform & 70/30 & 80/20 & 90/10 & 95/05 & 99/01 & zipf-1\\
\hline
Splay Tree & $2.3 \cdot {10}^{6}$ & $2.4 \cdot {10}^{6}$ & $2.5 \cdot {10}^{6}$ & $3.3 \cdot {10}^{6}$ & $4.1 \cdot {10}^{6}$ & $6.1 \cdot {10}^{6}$ & $5.2 \cdot {10}^{7}$ \\
\hline
SA2T & x$1.57$ & x$1.47$ & x$1.45$ & x$1.45$ & x$1.42$ & \cellcolor{blue!20}x$1.42$ & \cellcolor{blue!20}x$0.89$ \\
\hline
B-Tree & x$2.71$ & x$2.24$ & x$2.17$ & x$1.75$ & x$1.61$ & x$1.25$ & x$0.53$ \\
\hline
SABT & x$2.29$ & x$2.0$ & \cellcolor{blue!20}x$2.08$ & \cellcolor{blue!20}x$2.01$ & \cellcolor{blue!20}x$1.82$ & \cellcolor{blue!20}x$1.8$ & \cellcolor{blue!20}x$0.84$ \\
\hline
IST & x$2.1$ & x$1.68$ & x$1.64$ & x$1.31$ & x$1.1$ & x$0.89$ & x$0.2$ \\
\hline
SAIT & \cellcolor{blue!20}x$3.82$ & \cellcolor{blue!20}x$2.85$ & \cellcolor{blue!20}x$2.76$ & \cellcolor{blue!20}x$2.5$ & \cellcolor{blue!20}x$3.72$ & \cellcolor{blue!20}x$3.45$ & \cellcolor{blue!20}x$0.56$ \\
\hline
SALT & x$2.36$ & x$1.75$ & x$1.8$ & \cellcolor{blue!20}x$1.85$ & \cellcolor{blue!20}x$1.84$ & \cellcolor{blue!20}x$1.69$ & \cellcolor{blue!20}x$0.79$ \\
\hline
\end{tabular}
}
\end{table}

\begin{table}[!h]
\caption{lazy-delete mixed workloads, key set $|S| = 10^6$}
\label{table:lazy-delete-mixed-10^6}
\centering
\resizebox{\linewidth}{!}{%
\begin{tabular}{|c | c | c | c | c | c | c | c|}
\hline
 & uniform & 70/30 & 80/20 & 90/10 & 95/05 & 99/01 & zipf-1\\
\hline
Splay Tree & $1.1 \cdot {10}^{6}$ & $1 \cdot {10}^{6}$ & $1.1 \cdot {10}^{6}$ & $1.6 \cdot {10}^{6}$ & $2.1 \cdot {10}^{6}$ & $3.7 \cdot {10}^{6}$ & $5.4 \cdot {10}^{7}$ \\
\hline
SA2T & x$1.1$ & x$1.21$ & x$1.34$ & x$1.28$ & x$1.28$ & \cellcolor{blue!20}x$1.25$ & \cellcolor{blue!20}x$0.85$ \\
\hline
B-Tree & x$2.43$ & x$2.32$ & x$2.22$ & x$2.0$ & x$1.79$ & x$1.29$ & x$0.45$ \\
\hline
SABT & x$2.27$ & x$2.12$ & \cellcolor{blue!20}x$2.15$ & \cellcolor{blue!20}x$2.06$ & \cellcolor{blue!20}x$2.15$ & \cellcolor{blue!20}x$1.86$ & \cellcolor{blue!20}x$0.78$ \\
\hline
IST & x$1.58$ & x$1.57$ & x$1.68$ & x$1.22$ & x$1.12$ & x$0.92$ & x$0.2$ \\
\hline
SAIT & x$1.98$ & x$1.99$ & \cellcolor{blue!20}x$2.18$ & \cellcolor{blue!20}x$2.17$ & \cellcolor{blue!20}x$2.19$ & \cellcolor{blue!20}x$2.31$ & \cellcolor{blue!20}x$0.49$ \\
\hline
SALT & x$1.55$ & x$1.56$ & x$1.76$ & x$1.74$ & \cellcolor{blue!20}x$1.81$ & \cellcolor{blue!20}x$1.75$ & \cellcolor{blue!20}x$0.7$ \\
\hline
\end{tabular}
}
\end{table}

\begin{table}[H]
\caption{lazy-delete mixed workloads, key set $|S| = 10^7$}
\label{table:lazy-delete-mixed-10^7}
\centering
\resizebox{\linewidth}{!}{%
\begin{tabular}{|c | c | c | c | c | c | c | c|}
\hline
 & uniform & 70/30 & 80/20 & 90/10 & 95/05 & 99/01 & zipf-1\\
\hline
Splay Tree & $5.1 \cdot {10}^{5}$ & $4.9 \cdot {10}^{5}$ & $5 \cdot {10}^{5}$ & $6.2 \cdot {10}^{5}$ & $8.7 \cdot {10}^{5}$ & $1.8 \cdot {10}^{6}$ & $5.3 \cdot {10}^{7}$ \\
\hline
SA2T & x$0.97$ & x$1.05$ & x$1.05$ & x$0.86$ & x$0.75$ & x$0.56$ & \cellcolor{blue!20}x$0.62$ \\
\hline
B-Tree & x$2.76$ & x$2.66$ & x$2.56$ & x$2.24$ & x$1.87$ & x$1.11$ & x$0.37$ \\
\hline
SABT & x$2.28$ & x$2.14$ & x$2.13$ & x$2.1$ & x$1.63$ & \cellcolor{blue!20}x$1.2$ & \cellcolor{blue!20}x$0.62$ \\
\hline
IST & x$1.53$ & x$1.67$ & x$1.62$ & x$1.42$ & x$1.02$ & x$0.51$ & x$0.13$ \\
\hline
SAIT & x$1.75$ & x$1.54$ & x$1.5$ & x$1.62$ & x$1.36$ & \cellcolor{blue!20}x$1.51$ & \cellcolor{blue!20}x$0.36$ \\
\hline
SALT & x$1.58$ & x$1.47$ & x$1.33$ & x$1.44$ & x$1.26$ & x$0.77$ & \cellcolor{blue!20}x$0.51$ \\
\hline
\end{tabular}
}
\end{table}

Now, the results are similar to the results on read-only workloads~--- for GSATs each key remembers the previous number of accesses, so, when a key becomes unmarked the following operations to it will be fast.

\subsection{Analysis of distributions}
Now, let us prove that distributions used for the experiments, e.g., \texttt{zipf}, \texttt{90/10}, and \texttt{uniform}, are smooth for any $\alpha$. The following lemma proves it.

\begin{lemma}
\label{lem:bounded-distr}
Let $\mu$ be a probability density function for $[a, b]$ whose values on $[a, b]$ lies in $[x, y]$, where $x > 0$. Then, $\mu$ is smooth for any parameter $\alpha \in [\frac{1}{2}; 1)$.
\end{lemma}
\begin{proof}
At first, we fix a parameter $\alpha$, $n$ and $m = \lceil n^\alpha \rceil$. Then, consider any triple of reals $c_1$, $c_2$, $c_3$ such that $a \le c_1 < c_2 < c_3 \le b$. After denoting $c_3 - c_1$ as $len$, and $\frac{c_3 - c_1}{m}$ as $mlen$, we consider the integral from the definition of smoothness: $$\int_{c_2 - \frac{c_3 - c_1}{m}}^{c_2} \mu[c_1, c_3](t) \dt = \frac{\int_{c_2 - mlen}^{c_2} \mu(x) \dt}{\int_{c_1}^{c_3}\mu(t) \dt}  \le \frac{y \times mlen}{x \times len},$$ since $y$ is the maximum value of $\mu$, while $x$ is the minimum, thus,

$$\frac{y \times mlen}{x \times len} = \frac{y}{x \times m} \leq \frac{y}{x \times \sqrt{n}},$$
so, we can set $d = \frac{y}{x}$ in Definition~\ref{def:smooth} and the lemma is proved.
\end{proof}

All workloads from this paper have upper and lower bounds on the density function, thus, they are smooth.

\section{Conclusion}
\label{sec:conclusion}

The approach described in this paper can be applied to data structures with lazy rebuilding and result in self-adjusting versions. Applying to IST and B-Tree we get two self-adjusting data structures SAIT and SABT with the static-optimality property. Further, we proved that SAIT can get even better expected complexity.
Also, we provide a new data structure named Self-Adjusting Log Tree based on the observations from IST and B-Tree.
Then, we show how to design ``proper'' range queries on our data structures and how to make GSAT efficiently concurrent.
Finally, we run experiments to compare original trees and their self-adjusting counterparts~--- the more skew the workload has the better our data structures perform and sometimes outperform their vanilla versions, especially on read-only skewed workloads.

\bibliographystyle{plainurl}
\bibliography{references}

\appendix

\section{Auxiliary Lemmas and Theorems}

\begin{lemma}
\label{lem:gsat-reccurence}
If $T(m) = c \times \log_2(m) \times f(m) + (f(m) + 1) \times T\Big(\frac{m}{f(m) + 1}\Big)$ where $f(m)$ is \textit{sqrt-bounded} function, then $T(m) = O(m)$ for $m \in \mathbb{N}$.
\end{lemma}

\begin{proof}
For the proof, we assume that $m$ is at least $8$. Consider $\widetilde{T}(m) = \frac{T(m)}{m}$. Then $m\widetilde{T}(m) = c \times \log_2(m) \times f(m) + (f(m) + 1) \times \frac{m}{f(m) + 1} \times  \widetilde{T}\Big(\frac{m}{f(m) + 1}\Big)$, so, $\widetilde{T}(m) = \frac{c \times \log_2(m) \times f(m)}{m} + \widetilde{T}\Big(\frac{m}{f(m) + 1}\Big)$.
By sqrt-bounded definition we have $\frac{m}{f(m) + 1} \leq \frac{m}{2}$, thus, $\widetilde{T}(m) \leq \frac{c \times \log_2(m) \times f(m)}{m} + \widetilde{T}\Big(\frac{m}{2}\Big)$.
Now, we can unwrap the recurrence explicitly: $$\widetilde{T}(m) \leq \sum\limits_{i = 0}^{\log_2(m)} \frac{c \times \log_2(\frac{m}{2^i}) \times f(\frac{m}{2^i}) \times 2^i}{m} = \frac{c}{m} \times \Big( \sum\limits_{i = 0}^{\log_2(m)} f\Big(\frac{m}{2^i}\Big) \times \log_2\Big(\frac{m}{2^i}\Big) \times 2^i \Big) .$$

Let $M^*$ be the minimum value such that if $f(x) > \sqrt{x}$ holds $M^* \geq f(x)$.
Therefore, $$\widetilde{T}(m) \leq \frac{c}{m} \times \Big(\sum\limits_{i = 0}^{\log_2(m)} \Big( \sqrt{\frac{m}{2^i}} \times \log_2\Big(\frac{m}{2^i}\Big) \times 2^i \Big) +  M^* \times \sum\limits_{i = 0}^{\log_2(m)} \Big( \log_2\Big(\frac{m}{2^i}\Big) \times 2^i \Big) \Big).$$

Consider functions $$P(m) = \sum\limits_{i = 0}^{\log_2(m)}  p(i) \text{ where } p(x) = \sqrt{\frac{m}{2^x}} \times \log_2\Big(\frac{m}{2^x}\Big) \times 2^x$$ and $$Q(m) = \sum\limits_{i = 0}^{\log_2(m)} q(i) \text{ where } q(x) = \log_2\Big(\frac{m}{2^x}\Big) \times 2^x.$$

If we show that $P(m) = O(m)$, then $Q(m) = O(m)$, because $p(x) \geq q(x)$ if $x \geq 0$, and we get our target that $\widetilde{T}(m) = O(1)$ since $$\widetilde{T}(m) \leq \frac{c}{m} \times \Big( P(m) +  M^* \times Q(m) \Big) $$  $$= \frac{c}{m} \times \Big( O(m) +  M^* \times O(m) \Big) = c \times O(1) + c \times M^* \times O(1) = O(1),$$ because $M^*$ is a constant. Therefore, we get $T(m) = m \cdot \widetilde{T}(m) = m \cdot O(1) = O(m)$ and the lemma is proved.

Now, we prove that $P(m) = O(m)$.
By assuming that $m = 2^r$, we can write $\log m = r$ and  $p(x) = \sqrt{\frac{m}{2^x}} \times \log_2\Big(\frac{m}{2^x}\Big) \times 2^x = 2^{\frac{r - x}{2}} \times (r - x) \times 2^x = 2^{\frac{r + x}{2}} \times (r - x) = (\sqrt{2})^{r + x} \times (r - x)$. We want to find intervals where $p(x)$ is monotonous. For that we take a derivative $$ p'(x) = (\sqrt{2})^{r + x} \times \Big( \frac{\ln(2)}{2} (r - x) - 1 \Big) $$ and compare it with $0$. As we can see, $p'(x) = 0$ only if $x = r - 2\log_2{e} = x_0$. Therefore, the function $p(x)$ has only one point where the monotonicity changes. On the argument bigger than $x_0$, by substituting $x = r$, we can see that the derivative is negative. At the same time, on the argument less than $x_0$, by substituting $x = 0$, we get a positive value. So, $x_0$ is the point of maximum. 

From that we split $x$-s by $x_0$. If $x \geq r - 3$ we get $p(x) \leq p(x_0)$ since $2 \log_2(e) < 3$, hence, $$ P(m) =  \sum\limits_{i = 0}^{r} p(i) \leq \Big(\sum\limits_{i = 0}^{r - 3} p(i) \Big) + 3 \times f(x_0) \leq \Big( \int\limits_{0}^{r - 3} p(x) \dx \Big) + 4 \times f(x_0)$$ For the last inequality, we add $f(x_0)$ so $\sum\limits_{i = 0}^{r - 3} p(i) \leq (\int\limits_{0}^{r - 3} p(x) \dx ) + f(x_0)$.
Then, we calculate the indefinite integral: $$\int p(x) \dx = \int \Big( (\sqrt{2})^{r + x} \times (r - x) \Big) \dx = \frac{(\sqrt{2})^{r + x + 2}}{\ln^2(2)} \times (r\ln(2) - x\ln(2) + 2) + C.$$

Putting the integral we get $P(m) \leq ( \int\limits_{0}^{r - 3} p(x) \dx ) + 4 \times f(x_0) = \frac{(\sqrt{2})^{r + 2}}{\ln^2(2)} \times ( (\sqrt{2})^{r - 3} (3\ln(2) + 2) - r\ln(2) - 2 )$ $+ 4 \times ( \frac{2}{e} \times 2^r \log_2(e) ) \leq \frac{2^r}{\ln^2(2)} \times (3\ln(2) + 2) + \frac{8}{e} \times 2^r \log_2(e) = 2^r \times ( \frac{3}{\ln(2)} + \frac{2}{\ln^2{2}} + \frac{8\log_2(e)}{e})$ $= m c'$, where $c'$ is a constant, thus, $P(m) = O(m)$.
\end{proof}

\subsection{SALT}
\label{app:salt-proof}

\begin{theorem}
An ideal SALT requires $O(\min(n\log n, m))$ time to build, uses $O(n)$ memory, and has depth $O(\frac{\log m}{\log \log m})$.
\end{theorem}
\begin{proof}
SALT stores only elements and does not store an additional data structure, so, we immediately get the required bounds for the construction time and the memory consumption.

For SALT depth we have: $$d(m) \leq 1 + d\Big(\frac{m}{\log_2 m}\Big) \leq 2 + d\Big(\frac{m}{\log_2 m \times (\log_2 m - \log_2 \log_2 m)}\Big).$$ Let us find such $y$ that $\log_2 m - \log_2 \log_2 m \geq y > 0$.
Rewriting inequality through the exponentiation $\frac{m}{\log_2 m} \geq 2^y$ and assuming $\frac{m}{\log_2 m} \geq \sqrt{m} = 2^y$, we have $\log_2 m - \log_2 \log_2 m \geq y = \frac{1}{2} \times \log_2 m$, thus, by applying the inequality one additional time, $$d\Big(\frac{m}{\log_2 m \times (\log_2 m - \log_2 \log_2 m)}\Big) \leq d\Big(\frac{m \times 2}{\log_2 m \times \log_2 m}\Big)$$ $$\leq 1 + d\Big(\frac{m \times 2}{\log_{2}^2 m \times (\log m + 1 - 2\log_2 \log_2 m)}\Big).$$
Consider the function in the denominator above:  $$h_0(x) = \log m + x - (x + 1)\log_2 \log_2 m,$$ participating in such inequality: $$2^{h_0(x)} = \frac{m \cdot 2^x}{(\log_2 m)^{(x + 1)}} \geq \sqrt{m}.$$
While $h_0(x) \geq \frac{1}{2} \times \log_2 m$ holds
we get $$d\Big(\frac{m \times 2}{\log_{2}^2 m \times (\log m + 1 - 2\log_2 \log_2 m)}\Big) =  d\Big(\frac{m \times 2}{\log_{2}^2 m \times h_0(1)}\Big)\leq d\Big(\frac{m \times 2^2}{(\log_2 m)^3}\Big)$$ $$= d\Big( 2^{h_0(2)} \Big) \leq 1 + d\Big(\frac{m \times 2^2}{(\log_2 m)^3 \times (\log_2 m - 3 + 4 \log_2 \log_2 m)}\Big) \leq 1 + d\Big(\frac{m \times 2^3}{(\log_2 m)^4}\Big) $$ $$= 1 + d\Big( 2^{h_0(3)} \Big).$$

Thus, we can estimate the depth as: $$d(m) \leq 1 + t + d\Big( 2^{h_0(t)} \Big),$$ where $t$ is the first value with $$2^{h_0(t)} = \frac{m \cdot 2^{t}}{(\log_2 m)^{t + 1}} \leq \sqrt{m} \implies \log_2 m + t \leq \frac{1}{2} \times \log_2 m + (t + 1) \times \log_2 \log_2 m$$ $$\implies t \geq \frac{\frac{1}{2} \times \log_2 m - \log_2 \log_2 m}{\log_2 \log_2 m - 1} \geq \frac{\frac{1}{2} \times \log_2 m}{\log_2 \log_2 m - 1}.$$

So, $$d(m) \leq \Big( 1 + \frac{\frac{1}{2} \times \log_2 m}{\log_2 \log_2 m - 1} \Big)+ d(\sqrt{m}).$$

Next, we repeat the similar analysis but now for $d(\sqrt{m})$ instead of $d(m)$: $$d(\sqrt{m}) \leq 1 + d\Big(\frac{2 \times \sqrt{m}}{\log_2 m}\Big) \leq 2 + d\Big(\frac{2 \times \sqrt{m}}{\log_2 m \times (1 + \frac{1}{2} \times \log_2 m - \log_2 \log_2 m)}\Big).$$

Let $y$ be the value such that $(1 + \frac{1}{2} \times \log_2 m - \log_2 \log_2 m) \geq y > 0 \implies \frac{2 \sqrt{m}}{\log_2 m} \geq 2^y$.

If $\frac{2 \sqrt{m}}{\log_2 m} \geq m^{\frac{1}{4}} = 2^y$ holds, we have $y = \frac{1}{4} \times \log_2 m$, so $$d\Big(\frac{2 \times \sqrt{m}}{\log_2 m \times (1 + \frac{1}{2} \times \log_2 m - \log_2 \log_2 m)}\Big) \leq d\Big(\frac{2 \times \sqrt{m} \times 4}{(\log_2 m)^2}\Big)$$ $$\leq 1 + d\Big(\frac{2 \times \sqrt{m} \times 4}{(\log_2 m)^2 \times (1 + \frac{1}{2} \times \log_2 m + 2 - 2 \times \log_2 \log_2 m)}\Big).$$

Consider the function in the denominator above $$h_1(x) = 1 + \frac{1}{2} \times \log_2 m + 2x - (x + 1)\log_2 \log_2 m.$$ We have $h_1(x) \geq \frac{1}{4} \times \log_2 m$ until the inequality holds: $$2^{h_1(x)} = \frac{2\sqrt{m} \times 2^{2x}}{(\log_2 m)^{x + 1}} \geq m^{\frac{1}{4}},$$ therefore $$d\Big(\frac{2 \times \sqrt{m}}{(\log_2 m)^2 \times (1 + \frac{1}{2} \times \log_2 m + 2 - 2 \times \log_2 \log_2 m)}\Big) = d\Big(\frac{2 \times \sqrt{m} \times 4}{(\log_2 m)^2 \times h_1(1)}\Big)$$ $$\leq d\Big(\frac{2 \times \sqrt{m} \times 4^2}{(\log_2 m)^3 }\Big) = d\Big( 2^{h_1(2)}\Big)$$ $$\leq 1 + d\Big(\frac{2 \times \sqrt{m} \times 4^2}{(\log_2 m)^3 \times (1 + \frac{1}{2} \times \log_2 m + 4 - 3 \times \log_2 \log_2 m)}\Big)$$ $$= 1 + d\Big(\frac{2 \times \sqrt{m} \times 4^2}{(\log_2 m)^3 \times h_1(2)}\Big) \leq 1 + d\Big(\frac{2 \times \sqrt{m} \times 4^3}{(\log_2 m)^4 }\Big) = 1 + d\Big( 2^{h_1(3)}\Big).$$

So, $$d(\sqrt{m}) \leq 1 + t + d\Big(2^{h_1(t)}\Big),$$ where $t$ is the first value that $$\frac{2\sqrt{m} \times 2^{2t}}{(\log_2 m)^{t + 1}} \leq m^{\frac{1}{4}} \implies t \geq  \frac{1 + \frac{1}{4} \times \log_2 m - \log_2 \log_2 m}{\log_2 \log_2 m - 2} \geq \frac{1 + \frac{1}{4} \times \log_2 m}{\log_2 \log_2 m - 2} ,$$ hence we get a new estimate for the depth: $$d(m) \leq \Big(1 + \frac{\frac{1}{2} \times \log_2 m}{\log_2 \log_2 m - 1} \Big) + \Big(1 + \frac{1 + \frac{1}{4} \times \log_2 m}{\log_2 \log_2 m - 2} \Big) + d(m^{\frac{1}{4}}).$$

Using this approach, we performed two iterations moving from $d(m)$ to $d(m^{\frac{1}{2}})$ and from $d(m^{\frac{1}{2}})$ to $d(m^{\frac{1}{4}})$.
Now, let us consider $k$-th iteration: $$d(m^{\frac{1}{2^k}}) \leq 1 + d\Big(\frac{2^k \times m^{\frac{1}{2^k}}}{\log_2 m}\Big) \leq 2 + d\Big(\frac{2^k \times m^{\frac{1}{2^k}}}{\log_2 m \times (k + \frac{1}{2^k} \times \log_2 m - \log_2 \log_2 m))}\Big).$$

Consider the function $$h_k(x) = k + \frac{1}{2^k} \times \log m + (k + 1) x - (x + 1)\log_2 \log_2 m,$$ while the inequality holds $$ 2^{h_k(x)} = \frac{2^k \times m^{\frac{1}{2^k}} \times 2^{(k + 1) x}}{(\log_2 m)^{x + 1}} \geq m^{\frac{1}{2^{k + 1}}} \implies h_k(x) \geq \frac{1}{2^{k + 1}} \times \log_2 m.$$
So, $$d(m^{\frac{1}{2^k}}) \leq 1 + t + d\Big(\frac{2^k \times m^{\frac{1}{2^k}} \times 2^{t(k + 1)}}{(\log_2 m)^{t + 1}}\Big),$$
where $t$ is the first value with $$\frac{2^k \times m^{\frac{1}{2^k}} \times 2^{t(k + 1)}}{(\log_2 m)^{t + 1}} \leq m^{\frac{1}{2^{k + 1}}}$$ $$\implies t \geq \frac{k + \frac{1}{2^{k + 1}} \times \log_2 m - \log_2 \log_2 m}{\log_2 \log_2 m - (k + 1)} \geq \frac{k + \frac{1}{2^{k + 1}} \times \log_2 m}{\log_2 \log_2 m - (k + 1)},$$ which allows us to write down the recurrence explicitly, assuming that $m = 2^{2^r}$: $$d(m) \leq \sum\limits_{i = 0}^{\log_2 \log_2 m - 1} \Big( 1 + \frac{\frac{1}{2^{i + 1}} \times \log_2 m}{\log_2 \log_2 m - (i + 1)}\Big) = \sum\limits_{i = 0}^{r - 1} \Big(1 + \frac{2^{r - 1 - i}}{r - 1 - i} \Big) \leq r + \sum\limits_{i = 0}^{r - 1} \frac{2^i}{i}.$$
Then, using the inequality $\frac{2^i}{i} \leq \frac{2}{3} \times \frac{2^{i + 1}}{i + 1}$ and supposing that $r \geq 2$ we get: $$\sum\limits_{i = 0}^{r - 1} \frac{2^i}{i} \leq \frac{2^{r - 1}}{r - 1} \times \sum\limits_{i = 0}^{r - 1} \Big(\frac{2}{3}\Big)^i \leq 3 \cdot \frac{2^{r-1}}{r-1} \leq 3 \cdot \frac{2^r}{r},$$ so, the depth can be estimated as: $$d(m) \leq \log_2 \log_2 m + 3 \cdot \Big( \frac{\log_2 m}{\log_2 \log_2 m} \Big) = O\Big(\frac{\log_2 m}{\log_2 \log_2 m}\Big).$$ 
\end{proof}

\section{Pseudocode}
\label{app:pseudocode}
In this section, we present pseudocodes:
1) how to build an ideal GSAT (listing~\ref{lst:build-ideal-gsat});
2) and \texttt{get} operation (listing~\ref{lst:get-op}), \texttt{insert} operation (listing~\ref{lst:insert-op}), \texttt{delete} operation (listing~\ref{lst:delete-op}), and \texttt{Rebuild} operation (listing~\ref{lst:rebuild-op}).

\subsection{Ideal GSAT construction}
\label{app:pseudocode-ideal}

\begin{algorithm}[H]
\caption{Building an ideal GSAT for $n$ elements for the segment $[a, b]$.}
\label{lst:build-ideal-gsat}
\begin{lstlisting}[mathescape=true]
fun BuildIdeal(E[] elements, int[] ac, int n, float a, float b):
  pac = new int[n + 1]
  pac[0] = 0
  for i = 1 .. n:
    pac[i] = pac[i - 1] + ac[i]
  return Build(elements, ac, pac, a, b, 0, n)

fun Build(E[] elements, int[] ac, int[] pac, float a, float b, 
          int lt, int rt):
  if rt - lt $\le$ 0:
    return null
  m = pac[rt] - pac[lt]
  node = new Node(m, a, b)
  k = 0
  for i = 1 .. $\ceil{\operatorname{D}(m)}$:
    from, to = lt, rt
    while to - from > 1:
      m = $\floor{\frac{from + to}{2}}$
      if pac[m] - pac[lt] < $\ceil[\Big]{\frac{m}{\ceil{\operatorname{D}(m)} + 1}}$:
        from = m
      else:
        to = m
    node.rep[i] = elements[to]
    node.ac[i] = ac[to]
    node.children[i] = 
      Build(elements, ac, pac, a, elements[to].key, lt, to - 1)
    k = k + 1
    a = elements[to].key
    lt = to
    if lt == rt:
      break
  node.children[k + 1] = 
    Build(elements, ac, pac, a, b, lt, rt)
  return node

\end{lstlisting}
\end{algorithm}

\subsection{Operations Implementations}
\label{app:pseudocode-operations}

\begin{algorithm}[H]
\caption{\texttt{Get} operation}
\label{lst:get-op}
\begin{lstlisting}[mathescape=true]
fun get(Node $T$, K key):
  $T'$ = null
  result = null
  while $T$ $\neq$ null:
    $T$.c = $T$.c + 1
    if $T.c > \frac{T.im}{4}$ and $T'$ == null:
        $T'$ = $T$
    i = S($T$, key)
    if $T$.rep[i] == key:
      if not $T$.marked[i]:
        result = $T$.value[i]
      $T$.ac[i] = $T$.ac[i] + 1
      break
    $T$ = $T$.children[i]
  if $T' \neq$ null:
    $T'$ = Rebuild($T'$)
  return result
\end{lstlisting}
\end{algorithm}

\begin{algorithm}[H]
\caption{\texttt{Insert} operation}
\label{lst:insert-op}
\begin{lstlisting}[mathescape=true]
fun insert(Node $T$, K key, V value):
  $T'$ = null
  $P$, j = null, null
  while $T$ $\neq$ null:
    $T$.c = $T$.c + 1
    if $T.c > \frac{T.im}{4}$ and $T'$ == null:
        $T'$ = $T$
    i = S($T$, key)
    if $T$.rep[i] == key:
      if $T$.marked[i]:
        $T$.marked[i] = false
        $T$.value[i] = value
      $T$.ac[i] = $T$.ac[i] + 1
      break
    $P$, j = $T$, i
    $T$ = $T$.children[i]
  if $T$ == null:
    $P$.children[j] = new Node(key, value, ac=1)
  if $T' \neq$ null:
    $T'$ = Rebuild($T'$)
\end{lstlisting}
\end{algorithm}

\begin{algorithm}[H]
\caption{\texttt{Delete} operation}
\label{lst:delete-op}
\begin{lstlisting}[mathescape=true]
fun delete(Node $T$, K key):
  $T'$ = null
  while $T$ $\neq$ null:
    $T$.c = $T$.c + 1
    if $T.c > \frac{T.im}{4}$ and $T'$ == null:
        $T'$ = $T$
    i = S($T$, key)
    if $T$.rep[i] == key:
      if not $T$.marked[i]:
        $T$.marked[i] = true
      $T$.ac[i] = $T$.ac[i] + 1
      break
    $T$ = $T$.children[i]
  if $T' \neq$ null:
    $T'$ = Rebuild($T'$)
\end{lstlisting}
\end{algorithm}


\begin{algorithm}[H]
\caption{Rebuild operation}
\label{lst:rebuild-op}
\begin{lstlisting}[mathescape=true]
fun Rebuild(Node $T$):
  $T$ = BuildIdeal(Unmarked($T$))
  for $v \in T$: # iterate over all vertices in $T$
    $T_v$.c = 0
    $T_v$.im = $T_v$.m
  return $T$
\end{lstlisting}
\end{algorithm}

\end{document}